\documentclass[english, openright, 12pt]{book}%
\usepackage[utf8]{inputenc}
\usepackage[TS1,T1]{fontenc}

\usepackage{textcomp}
\usepackage{mathptmx}
\usepackage{xcolor}
\usepackage{pgfplots}
\usepackage{tikz}
\usepackage{graphicx}
\usepackage{pgfplots}
\usepackage{psfrag}
\usepackage{pdfpages}

\newsavebox{\test}

\savebox{\test}{\tikz \draw[green!20,line width=0.5pt,fill= green!20] plot [smooth cycle,samples=8,domain={1:8}] (\x*360/8+5*rnd:0.4cm+0.8cm*rnd);}
\usetikzlibrary{decorations.pathreplacing,decorations.pathmorphing,shapes}
\pgfmathsetseed{9}

\usepackage{amsmath,amssymb,amsfonts,mathrsfs,bm,cases}
\usepackage{cases}
\usepackage{subfig}
\usepackage{csquotes}

\usepackage{ulem}
\usepackage[a4paper,top=25mm, bottom=30mm, left=25mm, right=25mm]{geometry} 
\usepackage{caption}

\def \mycaptionmargin{1cm}

\usepackage{fancyhdr}
\usepackage{fancyheadings}

\usepackage{calligra}
\DeclareMathAlphabet{\mathcalligra}{T1}{calligra}{m}{n}
\DeclareFontShape{T1}{calligra}{m}{n}{<->s*[2]callig15}{}
\usepackage{frcursive}
\definecolor{Clarte}{rgb}{0,0,0}
\definecolor{Map2}{rgb}{0,0,0}
\definecolor{Comments}{rgb}{0.05,0.95,0.05}
\definecolor{Correc}{rgb}{0.01,0.71,0.17}
\def\cor{\color{Map2}}

\def\bla{\color{black}}
\newcommand{\nucol}{\textcolor{Clarte}{\nu}}
\def\micron{\,\hbox{\textmu}\textrm{m}}
\def\nm{\,\textrm{nm}}

\def\mtt{ IEEE Trans.\ Microwave Theory Tech.\ }

\def\josaa{ J.\ Opt.\ Soc.\ Am.\ A }
\def\josab{ J.\ Opt.\ Soc.\ Am.\ B }

\def\oc{ Opt.\ Commun.\ }

\newcommand{\bE}{\textbf{E}}

\newcommand{\bH}{\textbf{H}}
\newcommand{\bn}{\textbf{n}}
\newcommand{\bk}{\textbf{k}}
\newcommand{\bw}{\textbf{w}}
\newcommand{\bt}{\textbf{t}}
\newcommand{\be}{\textbf{e}}

\newcommand{\RE}{\Re {\textit{e}}}
\newcommand{\IM}{\Im {\textit{m}}}
\newcommand{\Lvect}{\mathscr{M}}
\newcommand{\Rvect}{\mathscr{R}}
\newcommand{\Svect}{\mathscr{S}}

\newcommand{\etal}{\textit{et al.}}
\newcommand{\e}{e}
\newcommand{\ic}{i}
\newcommand{\B}{\mathbf}
\newcommand{\tens}[1]{\smash{\underline{\underline{#1}}}}
\newcommand{\re}{\mathfrak{Re}}

\def\x{\mathbf{x}}
\def\y{\mathbf{y}}
\def\z{\mathbf{z}}

\def\rr{\mathbf{r}}

\def\build#1_#2^#3{\mathrel{\mathop{\kern 0pt#1}\limits_{#2}^{#3}}}

\def\bE{\ensuremath{\mathbf{E}}}
\def\br{\ensuremath{\boldsymbol{r}}}
\def\bH{\ensuremath{\mathbf{H}}}
\def\bk{\ensuremath{\mathbf{k}}}

\def\bF{\ensuremath{\boldsymbol{F}}}

\def\rot{\mathop{\rm rot}\nolimits}
\def\div{\mathop{\rm div}\nolimits}

\def\C{\mathbb C}

\def\grad{\mathop{\rm \mathbf{grad}}\nolimits}
\def\div{\mathop{\rm div}\nolimits}
\def\curl{\mathop{\rm  \mathbf{curl}}\nolimits}

\renewcommand{\RE}{\Re {\it{e}}}
\renewcommand{\IM}{\Im {\it{m}}}

\newcommand{\tenseps}{\underline{\underline{\varepsilon}}}
\newcommand{\tensmu}{\underline{\underline{\mu}}}
\newcommand{\tensdelta}{\underline{\underline{\delta}}}
\newcommand{\tensepst}{\underline{\underline{\tilde{\varepsilon}}}}
\newcommand{\tensmut}{\underline{\underline{\tilde{\mu}}}}
\newcommand{\tensdeltat}{\underline{\underline{\tilde{\delta}}}}

\newcommand{\tensxi}{\underline{\underline{\xi}}}

\def\grad{\mathop{\rm \mathbf{grad}}\nolimits}
\def\div{\mathop{\rm div}\nolimits}
\def\rot{\mathop{\rm \mathbf{curl}}\nolimits}

\def\e{\eta}

\def\x{\mathbf{x}}
\def\e{\mathbf{e}}

\newcommand{\Cnp}[2]%
{%
{#2 \choose #1}
}
\newcommand{\prop}{p}
\newcommand{\cprop}{c}
\def \logic{false}
\definecolor{bleu}{rgb}{0,0,.8}
\definecolor{bla}{rgb}{0,0,0}
\definecolor{red}{rgb}{.7,0,0}
\definecolor{vert}{rgb}{0,0.7,0}
\definecolor{purp}{rgb}{0.8,0.,0.9}
\def\todo{\color{bla}}
\def\toreadover{\color{bla}}
\def\bla{\color{bla}}

\def\newk{\color{bla}}

\begin{document}
\includepdf[pages=-]{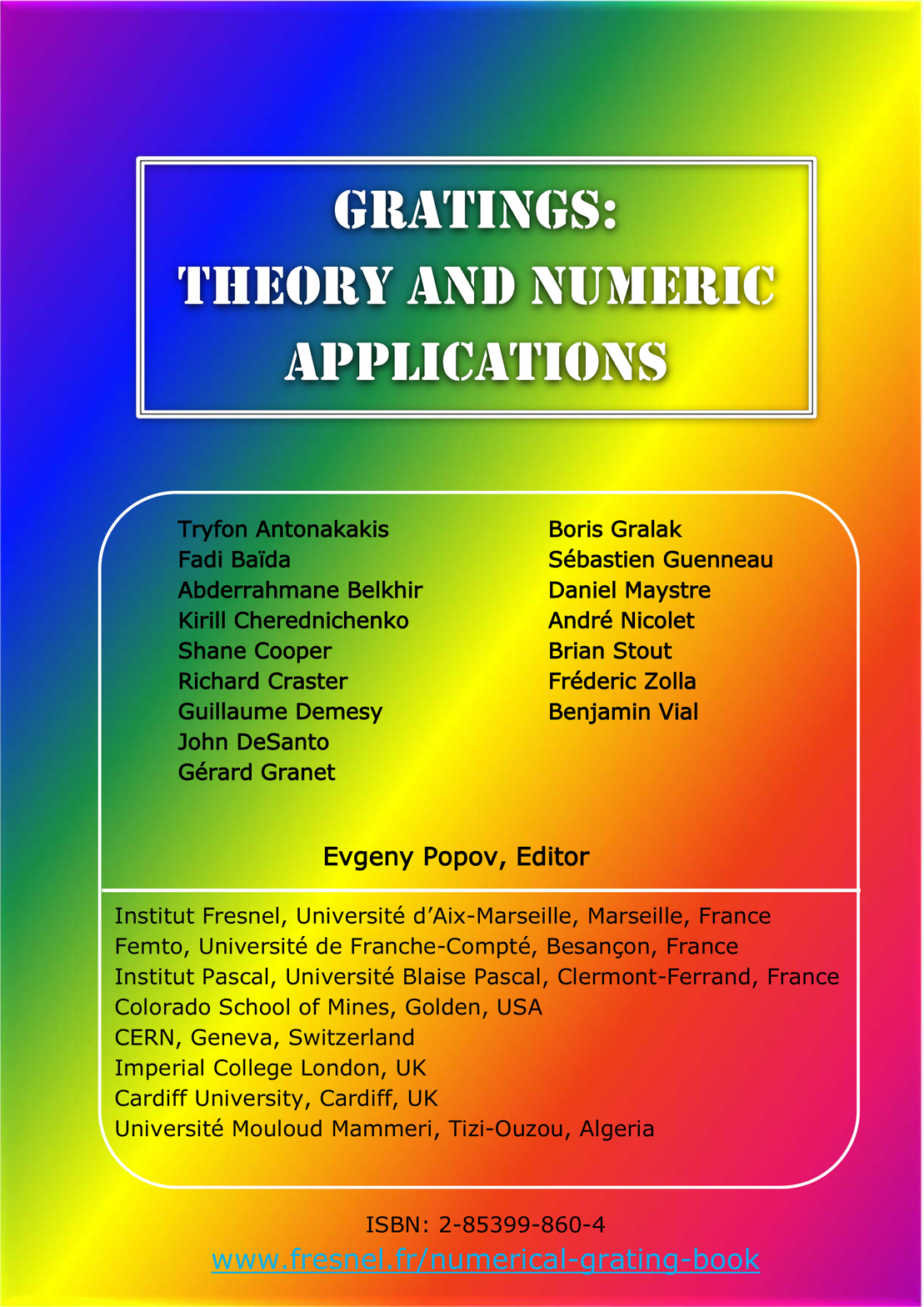}
\pagenumbering{arabic}
\newcommand{\numchapter}{5} 
\newcommand{\firstpage}{0} 
\setcounter{page}{\firstpage} 
\setcounter{chapter}{\numchapter} 

\let\cleardoublepage\clearpage
\renewcommand\contentsname{\normalsize{\hspace{12 pt}Table of Contents:}}
\tableofcontents
\thispagestyle{empty}
\newpage

\makeatletter
\renewcommand\section{\@startsection {section}{1}{\z@}%
{-3.5ex \@plus -1ex \@minus -.2ex}%
{2.3ex \@plus .2ex}%
{\reset@font\bfseries}}
\renewcommand\subsection{\@startsection {subsection}{1}{\z@}%
{-3.5ex \@plus -1ex \@minus -.2ex}%
{2.3ex \@plus .2ex}%
{\reset@font\it \bfseries}}
\renewcommand\subsubsection{\@startsection {subsubsection}{1}{\z@}%
{-3.5ex \@plus -1ex \@minus -.2ex}%
{2.3ex \@plus .2ex}%
{\reset@font\it \bfseries \underline}}

\thispagestyle{empty}
\begin{center}
	
{\fontsize{20}{20}\selectfont \textbf{
\hfill Chapter \numchapter\\[6mm]
Finite Element Method\\[10mm] 
}}
{\fontsize{14}{14}\selectfont Guillaume Dem{\'e}sy, Fr{\'e}d{\'e}ric Zolla, Andr{\'e}
  Nicolet, and Benjamin Vial \\[6mm] } 
{\fontsize{10}{10}\selectfont \textit{ 
Aix-Marseille Universit{\'e}, \'Ecole Centrale Marseille, Institut Fresnel,\\
13397 Marseille Cedex 20, France\\
}{\href{mailto:guillaume.demesy@fresnel.fr}{guillaume.demesy@fresnel.fr}}}
\end{center}
\pagestyle{fancyplain}
\renewcommand{\headrulewidth}{0.0pt} 
\lhead[\itshape{\small{\thechapter .\thepage}}]{\itshape{\small{G. Dem{\'e}sy \textit{et al.}: Finite Element Method}}}
\rhead[\itshape{\small{Gratings: Theory and Numeric Applications, 2012}}]{\itshape{\small{\thechapter .\thepage}}}
\cfoot{}

\section{Introduction}

\todo 


Finite element methods (FEM) represent a very general set of techniques to approximate
solutions of partial derivative equations. Their main advantage lies in their ability 
to handle arbitrary geometries via unstructured meshes of the domain of interest: The
discretization of oblic geometry edges is natively built in. Finite Element Methods have been
widely developed in many areas of physics and engineering: mechanics, thermodynamics\dots

But until the early 80's, two major drawbacks prevented them from being used 
in electromagnetic problems. On the one hand, existing nodal element basis 
did not satisfy the physical (dis)continuity of the vector fields components 
and lead to spurious solutions \cite{bossavit1990sme}. On the other hand, there 
was no proper way to truncate unbounded regions in open wave problems.

These two major limitations were both overcome in the early 80's: Vector elements 
have been developed by N\'ed\'elec \cite{nedelec1980mfe,bossavit1990solving},
and Perfectly Matched Layers (PMLs) were discovered by B\'erenger \cite{Berenger1}. 
Since then, it has been shown that PMLs could be described in the general framework 
of transformation optics
\cite{chew20073d,teixeira1998general,nicolet2008geometrical,agha2008use}.

All the mathematical and computational ingredients now exist and the goal of this 
chapter is to show how to combine them to implement a general 3D numerical scheme 
adapted to gratings using Finite Elements. In fact, we are now facing the physical 
difficulties inherent to the infinite spatial characteristics of the grating problem, 
whereas the computation domain has to be bounded in practice: 
(i) Both the superstrate and the substrate are infinite regions, 
(ii) there is an infinite number of periods and, last but not least,
(iii) the sources of the incident field (a plane wave) are located in the superstrate 
at an infinite distance from the grating.

In this chapter, the infinite extension of the superstrate and substrate is 
addressed using cartesian PMLs. In the framework of transformation optics, we demonstrate 
that B\'erenger's original PMLs can be extended to the challenging numerical cases of 
grazing incidence in order to deal with extreme oblic incidences or configurations 
near Wood's anomalies. The second issue of infinite number of period 
can be addressed via Bloch conditions. Finally, we are dealing with the distant plane wave sources 
through an equivalence of the diffraction problem with a radiation one whose 
sources are localized inside the diffractive element itself. The unknown field to 
be approximated using Finite Elements is a \textit{radiated field} 
with sources \textit{inside} the computation box and allows to retrieve easily the \textit{total field}
with the plane wave source.
\bla

\todo In a first section, we derive and implement this approach in the so-called 2D non-conical, or scalar, case.
We are dealing with the infinite issues rigorously in both TE and TM polarization cases. 
It results in a radiation problem with sources localized in the
diffractive element itself. We mathematically split the whole
problem into two parts. The first one consists in the classical
calculation of the \textit{total field} solution of a simple
interface. The second one amounts to looking for a \textit{radiated
field} with sources confined within the diffractive obstacles and
deduced from the first elementary problem. From this viewpoint, the
later \textit{radiated field} can be interpreted as an \textit{exact
perturbation} of the \textit{total field}. We show that our approach allows to
tackle some kind of anisotropy without increasing the computational
time or resource. Through a battery of examples, we illustrate 
its independence towards the geometry of the diffractive pattern.
Finally, we present an Adaptative PML able to tackle grazing incidences 
or configurations near Wood's anomaly.

In a second section, we extend this approach to the most general 
configuration of vector diffraction by crossed gratings embedded in 
arbitrary multilayered stack. The main
advantage of this method is, again, its complete independence towards the
shape of the diffractive element, whereas other methods often require heavy
adjustments depending on whether the geometry of the groove region presents 
oblique edges. This approach combined with the use of second order edge elements
allows us to retrieve the few numerical
academic examples found in the literature with an excellent accuracy. Furthermore, we provide a
new reference case combining major difficulties: A non
trivial toroidal geometry together with strong losses and a high
permittivity contrast. Finally, we discuss computation time and
convergence as a function of the mesh refinement
as well as the choice of the direct solver.

\bla

\section{Scalar diffraction by arbitrary mono-dimensional gratings : a Finite Element formulation}
\label{sec:2D}
\subsection{Set up of the problem and notations}
We denote by $\x$, $\y$ and $\z$, the unit vectors of the axes of an
orthogonal coordinate system $Oxyz$. We deal only with
time-harmonic fields; consequently, the electric and magnetic fields
are represented by the complex vector fields $\bE$ and $\bH$, with a
time dependance in $\exp(-i\, \omega\, t)$.

Besides, in this chapter, we assume that the tensor fields of relative
permittivity $\tenseps$ and relative permeability $\tensmu$ can be
written as follows:
\begin{equation}
\tenseps=%
\left( %
\begin{array}{ccc}
  \varepsilon_{xx} & \bar{\varepsilon}_{a} & 0 \\
  \varepsilon_{a} & \varepsilon_{yy} & 0 \\
  0 & 0 & \varepsilon_{zz}
\end{array}
\right) %
\quad \hbox{and} \quad
\tensmu=%
\left( %
\begin{array}{ccc}
  \mu_{xx} & \bar{\mu}_{a} & 0 \\
  \mu_{a} & \mu_{yy} & 0 \\
  0 & 0 & \mu_{zz}
\end{array}
\right) \; , %
\end{equation}
where $\varepsilon_{xx},\varepsilon_{a},\dots \mu_{zz}$ are possibly
complex valued functions of the two variables $x$ and $y$ and where
$\bar{\varepsilon}_{a}$ (resp. $\bar{\mu}_{a}$) represents the
conjugate complex of
$\varepsilon_{a}$ (resp. $\mu_{a}$). \textit{%
These kinds of materials are said to be $z$--anisotropic}. It is of
importance to note that with such tensor fields, lossy materials can
be studied (the lossless materials correspond to tensors with real
diagonal terms represented by Hermitian matrices) and that the
problem is invariant along the $z$--axis but the tensor fields can
vary continuously (gradient index gratings) or discontinuously (step
index gratings). Moreover we define $k_0:= \omega/c$.

The gratings that we are dealing with are made of three regions (See
Fig.~\ref{fig:Grating_Multi_Cells} ).
\begin{itemize}
\item%
\textit{The superstratum} ($y>h^g$) which is supposed to be
homogeneous, isotropic and lossless and characterized solely by its
relative permittivity $\varepsilon^+$ and its relative permeability
$\mu^+$ and we denote $k^+:=k_0\, \sqrt{\varepsilon^+ \mu^+}$%
\item%
\textit{The substratum} ($y<0$) which is supposed to be homogeneous
and isotropic and therefore characterized by its relative
permittivity $\varepsilon^-$ and its relative permeability
$\mu^-$ and we denote $k^-:=k_0\, \sqrt{\varepsilon^- \mu^-}$%
\item
\textit{The groove region} ($0<y<h^g$) which can be heterogeneous and
$z$--anisotropic and thus characterized by the two tensor fields
$\tenseps^g(x,y)$ and $\tensmu^g(x,y)$. It is worth noting that the
method does work irrespective of whether the
tensor fields are piecewise constant. The groove periodicity along
$x$--axis will be denoted $d$.
\end{itemize}
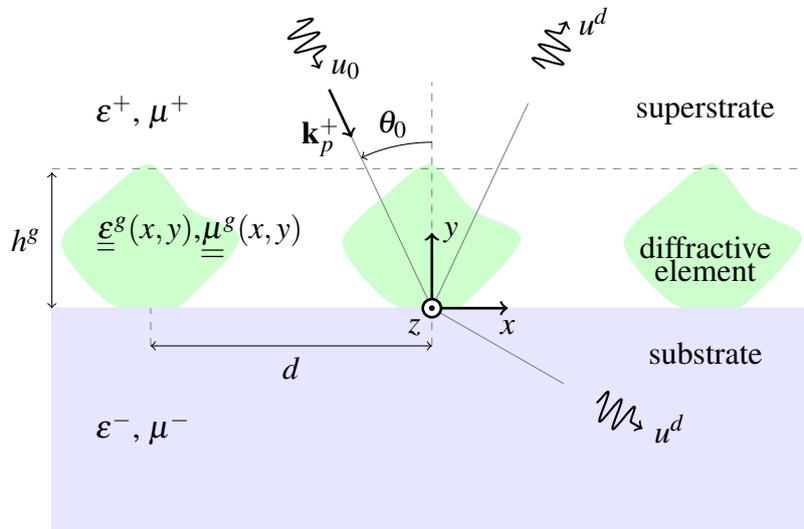
\begin{figure}[!ht] \centering
  \begin{tikzpicture}[
      media/.style={},
      wave/.style={
	  decorate,decoration={snake,post length=1.4mm,amplitude=2mm,
	  segment length=2mm},thick},
      interface/.style={
	  postaction={draw,decorate,decoration={border,angle=-45,
		      amplitude=0.3cm,segment length=2mm}}},
      ]
	 \draw (0,0.9) node {\usebox{\test}};
	 \draw (3.7,0.9) node {\usebox{\test}};
	 \draw (-3.7,0.9) node {\usebox{\test}};
      \fill[blue!10] (-5,-3) rectangle (5,0);
      \draw[blue!10,line width=0.5pt](-5,0)--(5,0);
      \draw[dashed,gray](-3.7,-0.5)--(-3.7,0);
      \draw[dashed,gray](0,-0.5)--(0,0);
	 \draw[dashed,gray](0,0)--(0,3);
	 \draw[<->](-5,0) --(-5,1.8)node[left]{} ;
	 \draw[](-5,0.9)node[left]{$h^g$} ;
	 \draw[dashed,gray](-5,1.85)--(5,1.85);
	 \draw[<->](-3.7,-.5) --(-1.85,-.5)node[below]{$d$}--(0,-.5) ;
      \draw(0.,-0.25)node[left]{$z$};
      \draw[<->,line width=1pt] (1,0) node[below]{$x$}-|(0,1) node[right]{$y$};
      \draw[->,wave] 
	  (115:4.2cm)--(115:3.5cm)node[right]{$u_0$};
      \draw[gray](0:0cm)--(115:3cm);
      \path (0,0)++(102:2.5cm)node{$\theta_0$};
      \draw[->](0,2.2)arc(90:115:2.2cm);
      \draw[->,line width=1pt]
	  (115:3.2cm)--(115:2.5cm)node[left]{$\mathbf{k}^+_\prop$};
      \draw[->,wave]
	  (-30:2.5cm)--(-30:3.2cm)node[right]{$u^d$};
      \draw[gray](0:0cm)--(-30:2cm);
      \draw[->,wave]
	  (65:3.5cm)--(65:4.2cm)node[right]{$u^d$};
      \draw[gray](0:0cm)--(65:3cm);
      \path[media] (3.6,2.6)  node {superstrate}
  (3.6,0.8)  node {diffractive}
  (3.6,0.5)  node {element}
		  (3.6,-.6) node {substrate};

      \path[media] (-3.8,2.6)  node {\normalsize$\varepsilon^+$, $\mu^+$}
  			    (-3.05,0.95)  node {\normalsize$\tenseps^g(x,y)$,$\tensmu^g(x,y)$}
		         (-3.8,-1.6) node {\normalsize$\varepsilon^-$, $\mu^-$};

      \filldraw[fill=white,line width=1pt](0,0)circle(.12cm);
      \filldraw[fill=black,line width=1pt](0,0)circle(.02cm);

  \end{tikzpicture}
\caption{Sketch and notations of the grating studied in this
section.}\label{fig:Grating_Multi_Cells}
\end{figure} 

This grating is illuminated by an incident plane wave of wave vector \newk
$\bk^+_\prop=\alpha\, \x - \beta^+ \, \y=k^+ \left(\sin \theta_0 \x -
 \cos \theta_0 \y \right)$, \bla whose electric field (TM case) (
 resp. magnetic field (TE case)) is linearly polarized along the $z$--axis:
\newk
\begin{equation}
\bE_e^0=\mathbf{A}_e^0 \, \exp(i \bk^+_\prop \cdot \br)\, \z \quad
(\hbox{resp. $\bH_m^0=\mathbf{A}_m^0 \, \exp(i \bk^+_\prop \cdot \br)\,
\z$}) \; ,
\end{equation}
\bla
where $\mathbf{A}_e^0$ (resp. $\mathbf{A}_m^0$) is an arbitrary
complex number and $\br =(x,y)^{\textrm{T}}$. \newk In this section, 
a plane wave is characterized by its wave-vector denoted  $\bk^{\{+,-\}}_{\{\prop,\cprop\}}$.
The subscript $\prop$ (resp. $\cprop$) stands for \enquote{propagative} (resp. \enquote{counter-propagative}). 
The superscript $+$ 
(resp. $-$) 
refers to the associated wavenumber $k^+$ (resp. $k^-$), and indicates that 
we are dealing with a plane wave propagating in the superstrate (resp. substrate).\bla
The magnetic (resp. electric) field derived from
$\bE_e^0$ (resp. $\bH_m^0$) is denoted $\bH_e^0$ (resp. $\bE_m^0$)
and the electromagnetic field associated with the incident field is
therefore denoted ($\bE^0,\bH^0$) which is equal to
($\bE_e^0,\bH_e^0$) (resp. ($\bE_m^0,\bH_m^0$)).

The diffraction problem that we address consists in
finding Maxwell equation solutions in harmonic regime
\textit{i.e.} the unique solution ($\bE,\bH$) of:
\begin{subequations}\label{eq:Maxwell}
\begin{numcases}{}
\curl\, \bE=i\, \omega\, \mu_0\, \tensmu \, \bH \label{eq:MaxwellrotE}\\
\curl\, \bH=-i\, \omega\, \varepsilon_0\, \tenseps \, \bE
\label{eq:MaxwellrotH}
\end{numcases}
\end{subequations}
such that the diffracted field $(\bE^{\textrm{d}},\bH^{\textrm{d}}):=(\bE-\bE^{0}_e,\bH-\bH^{0}_m)$ satisfies an \textit{Outgoing Waves
Condition} (O.W.C. \cite{petit1980electromagnetic}) and where $\bE$ and $\bH$
are quasi-periodic functions with respect to the $x$ coordinate.
\subsection{Theoretical developments of the method}
\subsubsection{Decoupling of fields and $z$--anisotropy}%
We assume that $\tensdelta(x,y)$ is a $z$--anisotropic tensor field
($\delta_{xz}=\delta_{yz}=\delta_{zx}=\delta_{zy}=0$). Moreover, the
left upper matrix extracted from $\tensdelta$  is denoted
$\tensdeltat$, namely:
\begin{equation}
\tensdeltat=%
\left( %
\begin{array}{cc}
  \delta_{xx} & \bar{\delta}_{a}  \\
  \delta_{a} & \delta_{yy}
\end{array}
\right) %
 \; . %
\end{equation}
For $z$--anisotropic materials, in a non-conical case, the problem
of diffraction can be split into two fundamental cases (TE case and
TM case). This property results from the following equality which
can be easily derived:%
\begin{equation}
- \rot \left ( {\tensdelta}^{-1} \rot \left( u \, \z \right)\right)=
\div \left ( {\tensdeltat^T}/\det(\tensdeltat) \nabla u \right) \z
\; ,
\end{equation}
where $u$ is a function which does not depend on the $z$ variable.
Relying on the previous equality, it appears that the problem of
diffraction in a non conical mounting amounts to looking for an
electric (resp. magnetic) field which is polarized along the
$z$--axis ; $\bE=e(x,y)\, \z$ (resp. $\bH=h(x,y)\, \z$). The
functions $e$ and $h$ are therefore solutions of similar
differential equations:
\begin{equation}\label{eq:defL}
\mathscr{L}_{\tensxi, \chi}(u):=\div\left( \tensxi \, \nabla u
\right) + k_0^2 \chi\, u =0
\end{equation}
with
\begin{equation}\label{eq:defxiTE}
u=e, \quad \tensxi= \tensmut^T/\det(\tensmut), \quad
\chi=\varepsilon_{zz} \; ,
\end{equation}
in the TM case and
\begin{equation}\label{eq:defxiTM}
u=h, \quad \tensxi= \tensepst^T/\det(\tensepst), \quad \chi=\mu_{zz}
\; ,
\end{equation}
in the TE case.

\subsubsection{Boiling down the diffraction problem to a radiation one}
In its initial form, the diffraction problem summed up by Eq.~(\ref{eq:defL}) is not well suited to the Finite Element Method. In
order to overcome this difficulty, we propose to split the unknown
function $u$ into a sum of two functions $u_1$ and $u_2^d$, the
first term being known as a closed form and the latter being a
solution of a problem of radiation \textit{whose sources are localized
within the obstacles}.

 We have assumed that outside the groove
region (cf. Fig.~\ref{fig:Grating_Multi_Cells}), the tensor field
$\tensxi$ and the function $\chi$ are constant and equal
respectively to $\tensxi^-$ and $\chi^-$ in the substratum ($y<0$)
and equal respectively to $\tensxi^+$ and $\chi^+$ in the
superstratum ($y>h^g$). Besides, for the sake of clarity, the
superstratum is supposed to be made of an isotropic and lossless
material and is therefore solely defined by its relative
permittivity $\varepsilon^+$ and its relative permeability $\mu^+$,
which leads to:
\begin{equation}\label{eq:defplusTE}
\tensxi^+= \frac{1}{\mu^+} \, \mathrm{Id}_2 \quad \hbox{and} \quad
\chi^+=\varepsilon^+ \quad \hbox{in TE case}
\end{equation}
or
\begin{equation}\label{eq:defplusTM}
\tensxi^+= \frac{1}{\varepsilon^+} \, \mathrm{Id}_2 \quad \hbox{and}
\quad \chi^+=\mu^+ \quad \hbox{in TM case,}
\end{equation}
where $\mathrm{Id}_2$ is the $2\times 2$ identity matrix. With such
notations, $\tensxi$ and $\chi$ are therefore defined as follows:
\begin{equation}\label{eq:def_xi_chi_global}
\tensxi(x,y):= \left \{
\begin{array}{lcc}
  \tensxi^+ & \hbox{for} & y>h^g \\
 \tensxi^g(x,y) & \hbox{for} & h^g>y>0\\
  \tensxi^- & \hbox{for} & y<0
\end{array}
\right .%
\; , \; \chi(x,y):= \left \{
\begin{array}{lcc}
  \chi^+ & \hbox{for} & y>h^g \\
 \chi^g(x,y) & \hbox{for} & h^g>y>0\\
  \chi^- & \hbox{for} & y<0 \; .
\end{array}
\right .
\end{equation}
It is now apropos to introduce an auxiliary tensor field $\tensxi_1$
and an auxiliary function $\chi_1$:
\begin{equation}
\tensxi_1(x,y):= \left \{
\begin{array}{ccc}
  \tensxi^+ & \hbox{for} & y>0 \\
  \tensxi^- & \hbox{for} & y<0
\end{array}
\right .%
\; , \; \chi_1(x,y):= \left \{
\begin{array}{ccc}
  \chi^+ & \hbox{for} & y>0 \\
  \chi^- & \hbox{for} & y<0 \; ,
\end{array}
\right .
\end{equation}
these quantities corresponding, of course, to a simple plane
interface. Besides, we introduce the constant tensor field
$\tensxi_0$ which is equal to $\tensxi^+$ everywhere and a constant
scalar field $\chi_0$ which is equal to $\chi^+$ everywhere.
Finally, we denote $u_0$ the function which equals the incident
field $u^{\mathrm{inc}}$ in the superstratum and vanishes elsewhere (see Fig.~\ref{fig:Grating_Multi_Cells}):
\begin{equation}
u_0(x,y):= \left \{
\begin{array}{ccc}
  u^{\mathrm{inc}} & \hbox{for} & y>h^g \\
  0 & \hbox{for} & y<h^g
\end{array}
\right .%
\end{equation}

We are now in a position to define more precisely the diffraction problem
that we are dealing with. The function $u$ is the unique solution of:
\begin{equation}\label{eq:defLOWC}
\mathscr{L}_{\tensxi, \chi}(u)=0 \,,\,\, \hbox{such that $u^d:=u-u_0$
satisfies an O.W.C.}
\end{equation}
In order to reduce this diffraction problem to a radiation
problem, an intermediate function is necessary. This function,
called $u_1$, is defined as the unique solution of the equation:
\begin{equation}\label{eq:defL1}
\mathscr{L}_{\tensxi_1, \chi_1}(u_1)=0 \,,\,\, \hbox{such that
$u_1^d:=u_1-u_0$ satisfies an O.W.C.}
\end{equation}
The function $u_1$ corresponds thus to \textit{an annex problem}
associated to a simple interface and can be solved in closed form
and \textit{from now on is considered as a known function}. As
written above, we need the function $u_2^d$ which is simply defined
as the difference between $u$ and $u_1$:
\begin{equation}\label{eq:u2d}
u_2^d:= u - u_1 =u^d-u_1^d\; .
\end{equation}
The presence of the superscript $d$ is, of course, not irrelevant:
As the difference of two diffracted fields, the O.W.C. of $u_2^d$ is
guaranteed (which is of prime importance when dealing with PML cf.
\ref{subsec:PML}). As a result, the Eq.~(\ref{eq:defLOWC}) becomes:
\begin{equation}\label{eq:radiating}
\mathscr{L}_{\tensxi, \chi}(u_2^d)=-\mathscr{L}_{\tensxi, \chi}(u_1)
\; ,
\end{equation}
where the right hand member is a scalar function which may be
interpreted as a \textit{known source term} $-\mathscr{S}_1(x,y)$
\textit{and the support of this source is localized only within the
groove region}. To prove it, all we have to do is to use Eq.~(\ref{eq:defL1}):
\begin{equation}
\mathscr{S}_1:=\mathscr{L}_{\tensxi,
\chi}(u_1)=\mathscr{L}_{\tensxi,
\chi}(u_1)-\underbrace{\mathscr{L}_{\tensxi_1,
\chi_1}(u_1)}_{=0}=\mathscr{L}_{\tensxi-\tensxi_1, \chi -
\chi_1}(u_1) \; .
\end{equation}
Now, let us point out that the tensor fields $\tensxi$ and
$\tensxi_1$ are identical outside the groove region and the same
holds for $\chi$ and $\chi_1$. The support of $\mathscr{S}_1$ is
thus localized within the groove region as expected. It remains to
compute more explicitly the source term $\mathscr{S}_1$. Making use
of the linearity of the operator $\mathscr{L}$ and the equality
$u_1=u_1^d+u_0$, the source term can be split into two terms
\begin{equation}\label{eq:defsource}
\mathscr{S}_1=\mathscr{S}_1^0+\mathscr{S}_1^d \; ,
\end{equation}
where
\begin{equation}\label{eq:defsource0}
\mathscr{S}_1^0=\mathscr{L}_{\tensxi-\tensxi_1, \chi - \chi_1}(u_0)
\end{equation}
and
\begin{equation}\label{eq:defsourced}
\mathscr{S}_1^d=\mathscr{L}_{\tensxi-\tensxi_1, \chi -
\chi_1}(u_1^d) \; .
\end{equation}
Now, bearing in mind that $u_0$ is nothing but a plane wave \newk $u_0= \,
\exp(i \bk^+_\prop \cdot \rr)$  (with $\bk^+_\prop = \alpha \x - \beta^+ \y$), \bla it
is sufficient to give $\nabla u_0= i \bk^+_\prop \, u_0$ for the weak
formulation associated with Eq. (\ref{eq:radiating}): \newk
\begin{equation}\label{eq:dev_source0}
\mathscr{S}_1^0= \left\{ i \div \left[ \left( \tensxi^+ -
\tensxi\right) \bk^+_\prop \, \exp(i \bk^+_\prop \cdot \rr) \right] + k_0^2\left(
\chi^+- \chi\right)\exp(i \bk^+_\prop \cdot \rr) \right\}\; .
\end{equation} \bla
The same holds for the term associated with the diffracted field. \newk
Since, in the superstrate, we have of course $u_1^d= \, \rho \exp(i \bk^+_\cprop \cdot \rr)$ with $\bk^+_\cprop = \alpha \x + \beta^+ \y$, 
\begin{equation}\label{eq:dev_source1}
\mathscr{S}_1^d= \rho \left\{ i \div \left[  \left( \tensxi^+ -
\tensxi\right) \bk^+_\cprop \, \exp(i \bk^+_\cprop \cdot \rr) \right] + k_0^2\left(
\chi^+- \chi\right)\exp(i \bk^+_\cprop \cdot \rr) \right\}\; ,
\end{equation} \bla
where $\rho$ is simply the complex reflection coefficient
associated with the simple interface:
\begin{equation}\label{eq:coeff_reflexion}
\rho = \frac{p^+-p^-}{p^+-p^-} \; \hbox{with}\;p^\pm\,=\, \left \{
\begin{array}{ccc}
    \;\beta^\pm\;\hbox{in the TM case}\\
    \\
    \;\frac{\beta^\pm}{\varepsilon^\pm}\;\hbox{in the TE case}\\
\end{array}
\right .
\end{equation}

\subsubsection{Quasi-periodicity and weak formulation}
The weak formulation follows the classical lines and is based on the
construction of a weighted residual of Eq.~(\ref{eq:defL}), which is
multiplied by the complex conjugate of a weight function $u'$ and
integrated by part to obtain:

\begin{eqnarray}
\mathscr{R}_{\tensxi, \chi}(u,u') = - \int_\Omega \left( \tensxi \,
\nabla u \right) \cdot \nabla \overline{u'} + k_0^2 \chi\, u  \;
\overline{u'} \; \textrm{d}\Omega + \int_{\partial \Omega} \overline{u'}
\left( \tensxi \, \nabla u \right) \cdot \mathbf{n}\;\textrm{d}S
\end{eqnarray}
The solution $u$ of the weak formulation can therefore be defined as
the element of the space $L^2(\curl,d,\alpha)$ of quasiperiodic functions
(i.e. such that $u(x,y)=u_\#(x,y)\,e^{i\alpha x}$ with
$u_\#(x,y)=u_\#(x+d,y)$, a $d$-periodic function) of $L^2(\curl)$ on
$\Omega$ such that: \begin{equation} \mathscr{R}_{\tensxi,
\chi}(u,u')=0 \; \; \forall u' \in L^2(\curl,d,\alpha). \end{equation} As
for the boundary term introduced by the integration by part,  it can
be classically set to zero by imposing Dirichlet conditions on a
part of the boundary (the value of $u$ is imposed and the weight
function $u'$ can be chosen equal to zero on this part of the
boundary) or by imposing homogeneous Neumann conditions $(\tensxi
\nabla u) \cdot \mathbf{n}=0$ on another part of the boundary (and
$u$ is therefore an unknown to be
determined on the boundary). 
\begin{figure}
\centering\includegraphics[width=0.7\textwidth
,draft=false]{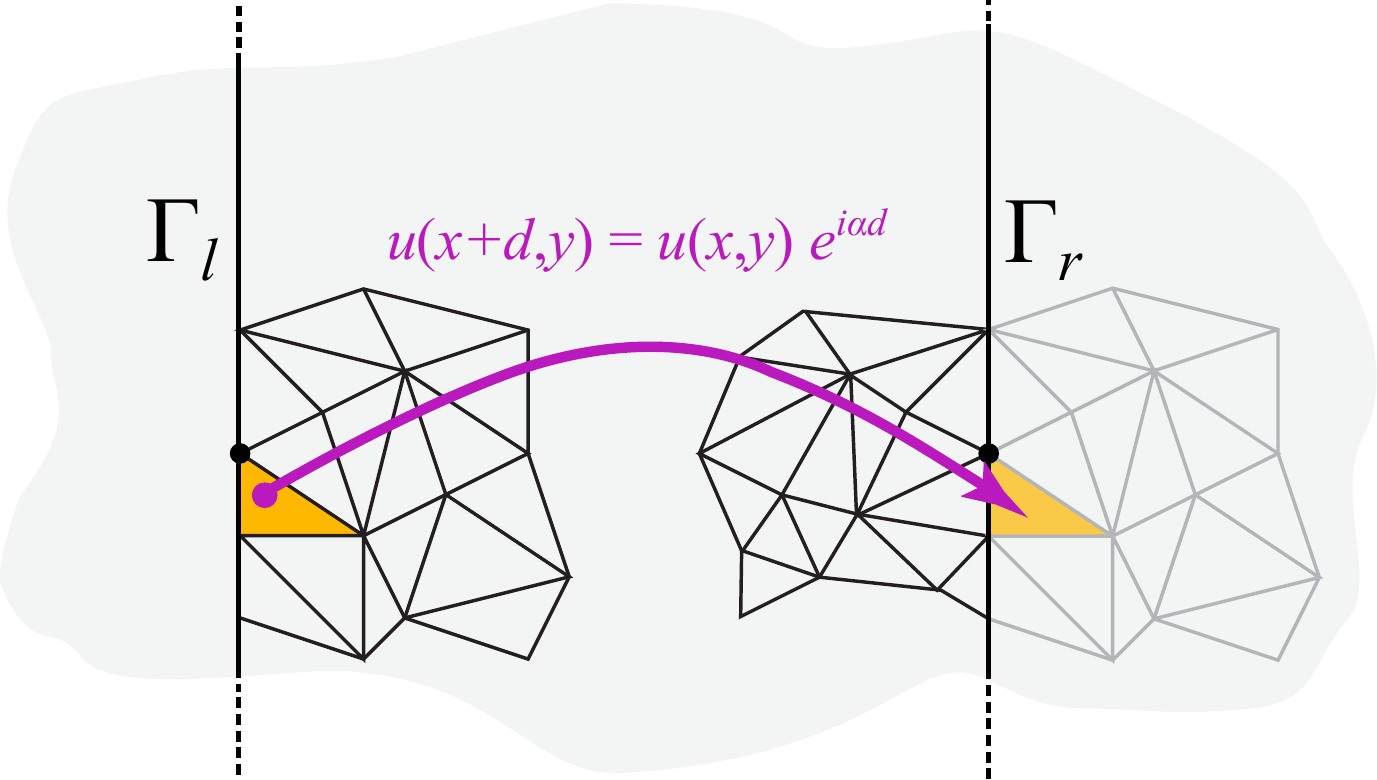} \caption{Quasi-periodicity of the field 
and sample of a $d$-periodic mesh.\label{fig:bloch} }
\end{figure}
A third possibility are the so-called
quasi-periodicity conditions of particular importance in the
modeling of gratings.

Denote by $\Gamma_l$ and $\Gamma_r$ the
lines parallel to the $y$--axis delimiting a cell of the grating (see Fig.~\ref{fig:bloch})
respectively from its left and right neighbor cell. Considering that both  $u$ and $u'$ are in $L^2(\curl,d,\alpha)$, the boundary term for $\Gamma_l \cup  \Gamma_r$ is

 $$\int_{\Gamma_l \cup \Gamma_r} \overline{u'} \left( \tensxi \, \nabla
u \right) \cdot \mathbf{n} \;\textrm{d}S= \int_{\Gamma_l \cup \Gamma_r}
\overline{u'_\#}e^{-i\alpha x} \left( \tensxi \, \nabla (u_\#e^{+i \alpha x})
\right) \cdot \mathbf{n} \;\textrm{d}S =$$
$$ \int_{\Gamma_l \cup \Gamma_r}
\overline{u'_\#} \left( \tensxi \,( \nabla u_\# +i \alpha u_\#
\mathbf{x})\right) \cdot \mathbf{n} \;\textrm{d}S = 0\,\, ,$$ because the integrand
$\overline{u'_\#} \left( \tensxi \,( \nabla u_\# +i \alpha u_\#
\mathbf{x})\right) \cdot \mathbf{n}$ is periodic along $x$ and the
normal $\mathbf{n}$ has opposite directions on $\Gamma_l$ and
$\Gamma_r$ so that the contributions of these two boundaries have
the same absolute value with opposite signs. The contribution of the
boundary terms vanishes therefore naturally in the case of
quasi-periodicity.

The finite element method is based on this weak formulation and both
the solution and the weight functions are classically chosen in a
discrete space made of linear or quadratic Lagrange elements, i.e.
piecewise first or second order two variable polynomial
interpolation built on a triangular mesh of the domain $\Omega$ (cf.
Fig.~\ref{subfig:PML_notations}). Dirichlet and Neumann conditions
may be used to truncate the PML domain in a region where the field
(transformed by the PML) is negligible. The quasi-periodic boundary
conditions are imposed by considering the $u$ as unknown on
$\Gamma_l$ (in a way similar to the homogeneous Neumann condition
case) while, on $\Gamma_r$, $u$ is forced equal to the value of the
corresponding point on $\Gamma_l$ (i.e. shifted by a quantity $-d$
along $x$) up to the factor $e^{i\alpha d}$. The practical implementation
in the finite element method is described in details in
\cite{zolla2012foundations,nicolet2004modelling}

\toreadover
\subsubsection{Perfectly Matched Layer for $z$--anisotropic
materials}\label{subsec:PML}%
 The main drawback encountered in
electromagnetism when tackling theory of gratings through the finite
element method is the non-decreasing behaviour of the propagating
modes in superstratum and substratum (if they are made of lossless
materials): The PML has been introduced by \cite{Berenger1} in
order to get round this obstacle. The computation of PML designed
for $z$--anisotropic gratings is the topic of what follows.

In the framework of transformation optics, a PML may be seen as a change of coordinate corresponding to a \textit{complex stretch} of the coordinate corresponding to the
direction along which the field must decay
\cite{Nicolet2,Lassas1,Lassas2}. 
Transformation optics have recently unified various techniques in
computational electromagnetics such as the treatment of open problems, helicoidal geometries or the design of invisibility cloaks (\cite{Nicolet2008}).
These apparently different problems share the same concept of geometrical transformation, leading to equivalent material properties. A very simple and practical rule
can be set up (\cite{zolla2012foundations}): when changing the coordinate system, all you have to do is to replace the initial materials properties $\tens{\varepsilon}$
 and $\tens{\mu}$ by equivalent material properties $\tens{\varepsilon}_s$ and $\tens{\mu}_s$ given by the following rule:

\begin{equation}
\tens{\varepsilon}_s=\B J^{-1}\,\tens{\varepsilon}\,\,\B J^{-\rm T}\,\mathrm{det}(\B J) \hspace{5pt}  \text{and~} \hspace{5pt} \tens{\mu}_s=\B J^{-1}\,\tens{\mu}\,\,\B J^{-\rm T}\,\mathrm{det}(\B J),
\label{equ_tranform}
\end{equation}
where $\B J$ is the Jacobian matrix of the coordinate transformation consisting of the partial derivatives of the new coordinates with respect to the original ones ($\B J^{-\rm T}$ is
the transposed of its inverse).\\
In this framework, the most natural way to define PMLs is to consider them as maps on a complex space $\mathbb{C}^3$, which coordinate change leads to equivalent permittivity and permeability tensors. We detail here the different coordinates used in this section.
\begin{itemize}
 \item $(x,y,z)$ are the cartesian original coordinates.
\item $(x_s,y_s,z_s)$ are the complex stretched coordinates. A suitable subspace $\Gamma \subset \mathbb{C}^3$ is chosen (with three real dimensions)
such that $(x_s,y_s,z_s)$ are the complex valued coordinates of a point on $\Gamma$ (e.g.\ $x=\re(x_s)$, $y=\re(y_s)$, $z=\re(z_s)$).
\item $(x_c,y_c,z_c)$ are three real coordinates corresponding to a real valued parametrization of $\Gamma \subset \mathbb{C}^3$.
\end{itemize}

We use rectangular PMLs (\cite{Nicolet2}) absorbing in the $y$-direction and
 we choose a diagonal matrix $\B J=\mathrm{diag}(1,s_y(y),1)$, where
 $s_y(y)$ is a complex-valued function of the real variable $y$, defined by:

\begin{equation}
 y_s(y)=\int_0^y s_y(y')\rm d y'.
\label{ys}
\end{equation}
The expression of the equivalent permittivity and permeability tensors are thus:
\begin{equation}
\tens{\varepsilon}_s=
 \left(
  \begin{array}{ c c c}
s_y\varepsilon_{xx} & \overline{\varepsilon_a} & 0 \\
 \varepsilon_a & s_y^{-1}\varepsilon_{yy} & 0 \\
  0 &  0 & s_y\varepsilon_{zz}
  \end{array} \right)
 \hspace{10pt}  \text{and~} \hspace{10pt}
\tens{\mu}_s=
 \left(
  \begin{array}{ c c c}
s_y\mu_{xx} & \overline{\mu_a} & 0 \\
 \mu_a & s_y^{-1}\mu_{yy} & 0 \\
  0 &  0 & s_y\mu_{zz}
  \end{array} \right).
\label{tenspml}
\end{equation}
\bla


\toreadover
Note that the equivalent medium has the same impedance than the original one as $\tens{\varepsilon}$ an $\tens{\mu}$ are transformed in the same way,
 which guarantees that the PML is perfectly reflectionless.\\
Now, let us define the so-called substituted field $\bF_s=(\B E_s, \B H_s)$, solution of Eqs.~(\ref{eq:Maxwell}) with $\tens{\xi}=\tens{\xi}_s$ and $\chi=\chi_s$.
It turns out that $\bF_s$ equals the field $\bF$ in the region $y^b<y<y^t$ (with $y^b=-h^-$ and $y^t=h^g+h^+$, see Fig.~\ref{subfig:PML_notations}), provided that $s_y(y)=1$ in this region.
The main feature of this latest field $\bF_s$ is the
remarkable correspondence with the first field $\bF$ ; whatever the
function $s_y$ provided that it equals $1$ for $y^t <y<y^b$, 
the two fields $\bF$ and $\bF_s$
are identical in the region $y^t <y<y^b$\cite{agha2008use}. In
other words, the PML is completely reflection-less. In addition, for
complex valued functions $s_{y}$ ($\IM \{s_{y} \}$ strictly positive
in PML), the field $\bF_s$ converges exponentially towards zero
(as $y$ tends to $\pm \infty$, cf. Fig.~\ref{subfig:PML_field} and
\ref{subfig:PML_log_field}) although its physical counterpart $\bF$
does not. Note that in Fig.~\ref{subfig:PML_log_field}, the value of 
the computed radiated field $u_2^d$ on
each extreme boundary of the PMLs is at least $10^{-8}$ weaker than
in the region of interest. As a consequence, $\bF_s$ is of finite energy and for this
substituted field a weak formulation can be easily derived which is
essential when dealing with Finite Element Method.

Still remains to give a suitable function $s_y$. Let us consider the complex 
coordinate mapping $y(y_c)$, which is simply defined as the derivative of the stretching coefficient $s_y(y)$ with respect to $y_c$.
With simple stretching functions, we can obtain a reliable criterion upon proper fields decay. A classical choice is:
\begin{equation}
 s_y(y)=
\begin{cases}
 \zeta^- & \mbox{if }y<y^b\\
1 & \mbox{if }y^b<y<y^t\\
 \zeta^+ & \mbox{if }y>y^t\\
\end{cases}
\end{equation}
where $\zeta^{\pm}=\zeta^{',\pm}+i\zeta^{'',\pm}$ are complex constants with $\zeta^{'',\pm}>0$.\\
In that case, the complex valued function $y(y_c)$ defined by Eq.~(\ref{ys}) is explicitly given by:
\begin{equation}
 y(y_c)=
\begin{cases}
 y^b+\zeta^-(y_c-y^b)& \mbox{if }y_c<y^b\\
y_c & \mbox{if }y^b<y_c<y^t\\
 y^t+\zeta^+(y_c-y^t) & \mbox{if }y_c>y^t\\
\end{cases},
\end{equation}\\


\setlength{\captionmargin}{-.2cm}
\begin{figure}[!ht]
    \begin{center}
    \subfloat[Computational domain $\Omega$ and its five constituent regions.] {\includegraphics[width=2.6 cm,draft=\logic]{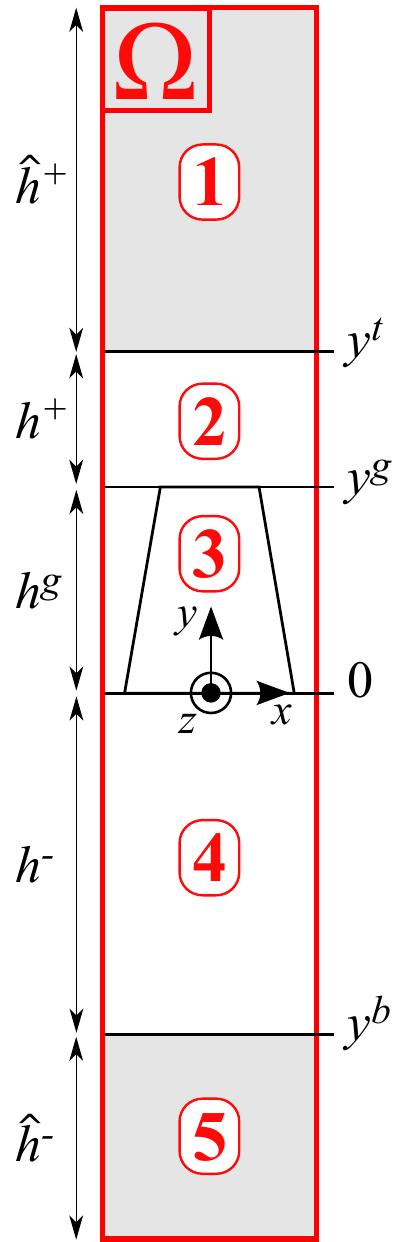}\label{subfig:PML_notations}}
    \qquad \qquad
    \subfloat[Coarse triangle meshing of the cell $\Omega$. Maximum element side size: $\lambda/(2\sqrt{\varepsilon})$] {\includegraphics[width=2.6 cm,draft=\logic]{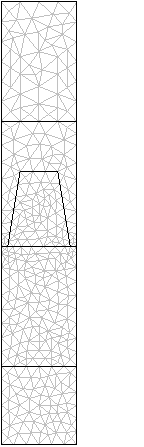}\label{subfig:PML_mesh}}
    \qquad
    \subfloat[Radiated field: $\RE\{u_2^d\}$ in $V/m$ ] {\includegraphics[width=2.6 cm,draft=\logic]{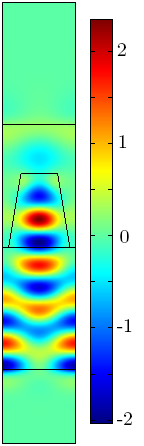}\label{subfig:PML_field}}
    \qquad
    \subfloat[Radiated field: $\log(|u_2^d|)$]{\includegraphics[width=2.6 cm,draft=\logic]{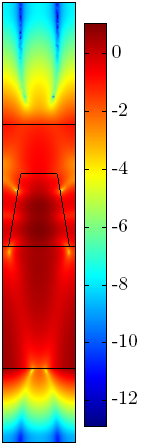}\label{subfig:PML_log_field}}
\setlength{\captionmargin}{\mycaptionmargin}
\caption{Example of computation of the radiated field $u_2^d$ (TM case).} \label{fig:PML}
    \end{center}
\end{figure}
\setlength{\captionmargin}{\mycaptionmargin}

\bla


\newk
Finally, let us consider a propagating plane wave in the substratum
$u_n(x,y):=\exp(i(\alpha x - \beta^-_{n} y))$. Its expression can be rewritten
as a function of the  stretched coordinates in the PML as follows:
\begin{equation}\label{eq:wave_in_PML}
u^{\mathrm{sc}}_n(x_c,y_c):=u_n(x(x_c),y(y_c))=e^{i\alpha x_c} e^{-i
\beta^-_{n}(y^b +\zeta^-(y_c-y^b))}
\end{equation}
The behavior of this latest function along the $y_c$ direction is
governed by the function $U^{\mathrm{sc}}(y_c):=e^{-i
\beta^-_{n}\, \zeta^- y_c}$. Letting
$\beta'^{,-}_{n}:=\RE\{\beta^-_{n}\}$, %
$\beta''^{,-}_{n}:=\IM \{ \beta^-_{n}\}$, %
$\zeta'^{,-}:=\RE \{\zeta^-\}$ and %
$\zeta''^{,-}:=\IM\{\zeta^-\}$,  the non-oscillating part of the
function $U^{\mathrm{sc}}(y_c)$ is given by
$\exp\left((\beta'^{,-}_{n} \, \zeta''^{,-}\, + \beta''^{,-}_{n} \,
\zeta'^{,-})y_c\right)$. Keeping in mind that $\beta'^{,-}_{n} $
and/or $\beta''^{,-}_{n} $ are positive numbers, the function
$U^{\mathrm{sc}}$ decreases exponentially towards zero as $y_c$
tends to $-\infty$ (Fig.~\ref{subfig:PML_log_field}) provided that
$\zeta^-$ belongs to $\C^+:=\{z \in \C, \RE\{z\}>0, \; \hbox{and} \;
\IM\{z\}>0\}$. In the same way, it can be shown that $\zeta^+$
belongs to $\C^+$.

\toreadover 
Let us conclude this section with two important remarks:
\begin{enumerate}
\item \textbf{Practical choice of PML parameters.} As for the complex 
stretch parameters, setting $\zeta^\pm=1+i$ is usually a safe choice. For computational needs,
the PML has to be truncated and the other constitutive parameter 
of the PML is its thickness $\hat{h}$ 
(see Fig.~\ref{subfig:PML_notations}). Setting $\hat{h}^\pm=\lambda_0/\sqrt{\varepsilon^\pm}$ 
leads to a PML thick enough to \enquote{absorb} all incident radiation. 
These specific values will be used in the sequel, unless otherwise specified.
\item \textbf{Special cases.} The reader will notice that a configuration 
where $\beta'^{,-}_{n}$ is a very weak positive number compared to $k_0$ 
with $\beta''^{,-}_{n}$ (this is precisely the case 
of a plane wave at grazing incidence on the bottom PML) 
\textbf{leads to a very slow exponential decay} of $U^{\mathrm{sc}}$. In such a case, close to so-called 
Wood's anomalies or at extreme grazing incidences, classical PML fail.
We will address this tricky situation extensively in Section~\ref{sec:adaptative_pml}.
\end{enumerate}
\bla

\subsubsection{Synthesis of the method}
In order to give a general view of the method, all information is
collected here that is necessary to set up the practical Finite
Element Model. First of all, the computation domain $\Omega$ (cf.
Fig.~\ref{subfig:PML_notations}) corresponds to a truncated cell of
the grating which is a finite rectangle divided into five horizontal
layers. These layers are respectively from top to bottom upper PML,
the superstratum, the groove region, the substratum, and the lower
PML. The unknown field is the scalar function $u_2^d$ defined in Eq.~(\ref{eq:u2d}). Its finite element approximation is based on the
second Lagrange elements built on a triangle meshing of the cell
(cf. Fig.~\ref{subfig:PML_mesh}). A complex algebraic system of
linear equations is constructed via the Galerkin weighted residual
method, \textit{i.e.} the set of weight functions $u'$ is chosen as
the set of shape functions of interpolation on the mesh
\cite{zolla2012foundations}.

\begin{itemize}
\item
In  region $1$ (upper PML, see Fig.~\ref{subfig:PML_notations}),
\begin{equation}
\mathscr{R}_{\tensxi_s^+, \chi_s^+}(u_2^d,u') = 0 \; ,
\end{equation}
with $\tensxi_s^+$ and $\chi_s^+$ depending on the equivalent
anisotropic properties of the PML given by Eq.~(\ref{eq:defxiTE}),
Eq.~(\ref{eq:defxiTM}) and Eqs.~(\ref{tenspml}).

\item
In region $2$ (superstratum),
\begin{equation}
\mathscr{R}_{\tensxi^+, \chi^+}(u_2^d,u') = 0 \; ,
\end{equation}
with $\tensxi^+$ and $\chi^+$ depending on the homogeneous isotropic
properties of the superstratum given by Eq.~(\ref{eq:defxiTE}), Eq.~(\ref{eq:defxiTM}), Eq.~(\ref{eq:defplusTE}) and Eq.~(\ref{eq:defplusTM}).

\item
In region $3$ (groove region),
\begin{equation}
\mathscr{R}_{\tensxi^g, \chi^g}(u_2^d,u') = -\mathscr{R}_{\tensxi^g,
\chi^g}(\mathscr{S}_1,u')\; ,
\end{equation}
with $\tensxi^g$ and $\chi^g$ depending on the heterogeneous
possibly anisotropic properties given by Eq.~(\ref{eq:defxiTE}), Eq.~(\ref{eq:defxiTM}), Eq.~(\ref{eq:def_xi_chi_global}) and
$\mathscr{S}_1$ given by Eq.~(\ref{eq:defsource}) , Eq.~(\ref{eq:dev_source0}), Eq.~(\ref{eq:dev_source1}) and Eq.~(\ref{eq:coeff_reflexion}).

\item
In region $4$ (substratum),
\begin{equation}
\mathscr{R}_{\tensxi^-, \chi^-}(u_2^d,u') = 0 \; ,
\end{equation}
with $\tensxi^-$ and $\chi^-$ depending on the homogeneous isotropic
properties of the substratum given by Eq.~(\ref{eq:defxiTE}), Eq.~(\ref{eq:defxiTM}), Eq.~(\ref{eq:defplusTE}) and Eq.~(\ref{eq:defplusTM}).

\item
In  region $5$ (lower PML),
\begin{equation}
\mathscr{R}_{\tensxi_s^-, \chi_s^-}(u_2^d,u') = 0 \; ,
\end{equation}
with $\tensxi_s^-$ and $\chi_s^-$ depending on the equivalent
anisotropic properties of the PML given by Eq.~(\ref{eq:defxiTE}),
Eq.~(\ref{eq:defxiTM}) and Eqs.~(\ref{tenspml}).
\end{itemize}
\subsubsection{Energy balance: Diffraction efficiencies and absorption}\label{section:diff_eff2D}
The rough result of the FEM calculation is the complex \textit{radiated} field 
$u_2^d$. Using Eq.~(\ref{eq:u2d}), it is straightforward to obtain the 
complex \textit{diffracted} field $u^d$ solution of Eq.~(\ref{eq:defLOWC}) 
at each point of the bounded domain.
We deduce from $u^d$ the diffraction
efficiencies with the following method. The superscripts $ ^+$
(resp. $ ^-$) correspond to quantities defined in the superstratum
(resp. substratum) as previously.

On the one hand, since $u^d$ is quasi-periodic along the $x$--axis ,
it can be expanded as a Rayleigh expansion (see for instance
\cite{petit1980electromagnetic}):

\begin{equation}\label{eq:devfourier}
\hbox{for}\;y<0\;\hbox{and}\;y>h^g,\;u^{d}(x,y) = \sum_{n \in
\mathbb{Z}}\,u^d_n(y)\,e^{i\alpha_n x}
\end{equation}
where
\begin{equation}\label{eq:coefffourier1}
u^d_n(y)= \frac{1}{d}\int_{-d/2}^{d/2}u^d(x,y)e^{-i\alpha_n x}\,\textrm{d}x
 \; \hbox{with} \; \alpha_n = \alpha+\frac{2\pi}{d}n
\end{equation}

On the other hand, introducing Eq.~(\ref{eq:devfourier}) into Eq.~(\ref{eq:defL}) leads to the Rayleigh coefficients:
\newk
\begin{equation}\label{eq:coefffourier2a}
u^d_n(y)= \left \{
\begin{array}{ccc}
  u_n^+(y)=r_n\,e^{\,i\beta_n^+ y} + a_n\,e^{\,-i\beta_n^+ y} & \hbox{for} & y>h^g \\
  \\
  u_n^-(y)=t_n\,e^{\,-i\beta_n^- y} + b_n\,e^{\,i\beta_n^- y} & \hbox{for} & y<0 \\
\end{array}
\right . \; \hbox{with} \; \beta_{n}^{\pm^2} = k^{\pm^2} -
\alpha_n^2
\end{equation} 
\bla
For a temporal dependance in \newk $e^{-i\omega t}$, \bla the O.W.C. imposes
$a_n=b_n=0$. Combining Eq.~(\ref{eq:coefffourier1}) and~(\ref{eq:coefffourier2a}) at a fixed $y_0$ altitude leads to:
\newk
\begin{equation}\label{eq:AnBn}
\left \{
\begin{array}{ccc}
        r_n = \frac{1}{d} \displaystyle \int_{-d/2}^{d/2}u^d(x,y_0)\,e^{-i(\alpha_n x+\beta_n^+ y_0)}\,\textrm{d}x & \hbox{for} & y_0>h^g\\
    \\
        t_n = \frac{1}{d} \displaystyle \int_{-d/2}^{d/2}u^d(x,y_0)\,e^{-i(\alpha_n x-\beta_n^- y_0)}\,\textrm{d}x & \hbox{for} & y_0<0\\
\end{array}
\right .
\end{equation}
\bla
We extract these two coefficients by trapezoidal numerical
integration along $x$ from a cutting of the previously calculated
field map at $y_0$. It is well known that the mere trapezoidal
integration method is very efficient for smooth and periodic
functions (integration on one period) \cite{helluy}. Now the
restriction on a horizontal straight line crossing the whole cell in
homogeneous media (substratum and superstratum) is of $C^{\infty}$
class. From a numerical point of view, it appears that the
interpolated approximation of the unknown function, namely $u_2^d$
preserves the good behaviour of the numerical computation of these
integrals. From this we immediately deduce the reflected and
transmitted diffracted efficiencies of propagative orders ($T_n$ and
$R_n$) defined by:
\begin{equation}\label{eq:diffefficiency}
\left \{
\begin{array}{ccc}
    R_n:=\,r_n\,\overline{r_n}\,\frac{\beta_n^+}{\beta^+} & \hbox{for} & y_0>h^g\\
    \\
    T_n:=\,t_n\,\overline{t_n}\,\frac{\beta_n^-}{\beta^-}\,\frac{\gamma^+}{\gamma^-} & \hbox{for} & y_0<0\\
\end{array}
\right . \hbox{with}\;\gamma^\pm\,=\, \left \{
\begin{array}{ccc}
    1\;\hbox{in the TM case}\\
    \\
    \;\varepsilon^\pm\;\hbox{in the TE case}\\
\end{array}
\right .
\end{equation}
This calculation is performed at several different $y_0$ altitudes in
the superstratum and the substratum, and the mean value found for each
propagative transmitted or reflected diffraction order is presented
in the numerical experiments of the following section.
\toreadover

Normalized losses $Q$ can be obtained according 
to Poynting's theorem through the straightforward computation of the following ratio:
\begin{equation}\label{eq:losses2D}
    Q := \frac{\displaystyle \int_{S}\omega\,\varepsilon_0\,\IM(\varepsilon^{g'})\,\bE\cdot\overline{\bE}\,\mathrm{d}s}
             {\displaystyle \int_{L} \RE\{\bE_0\times\overline{\bH_0}\}\cdot \textbf{n}\,\mathrm{d}l}
    \,,
\end{equation}
The numerator in Eq.~(\ref{eq:losses2D}) clarifies losses in Watts by
period of the considered grating and are computed by
integrating the Joule effect losses density over the surface $S$ of
the lossy element. The denominator normalizes these losses to the
incident power, \textit{i.e.} the time-averaged incident Poynting
vector flux across one period (a straight line $L$ of length $d$ in 
the superstrate parallel to $Ox$, whose normal oriented along decreasing values of $y$ is denoted $\bn$).

Finally, combining Eqs.~(\ref{eq:diffefficiency}) and Eq.~(\ref{eq:losses2D}), 
a self consistency check of the whole numerical scheme
consists in comparing the quantity $B$:
\begin{equation}\label{eq:balance2D}
    	B:=\displaystyle \sum_n T_n +\sum_m R_m + Q
\end{equation}
to unity. In Eq.~(\ref{eq:balance2D}), the summation indexed by $n$ (resp. $m$) 
corresponds to the sum over the efficiencies of all transmitted (resp. reflected) 
propagative diffraction orders in the substrate (resp. superstrate).
We give interpretations and concrete examples of such numerical energy 
balances over non trivial grating profiles in 
sections~\ref{sec:aniso_example}~and~\ref{sec:marsu}.

\bla


\subsection{Numerical experiments}
\subsubsection{Numerical validation of the method}
\begin{figure}[!ht]
\center{
  \includegraphics[width=.7 \textwidth,draft=\logic]{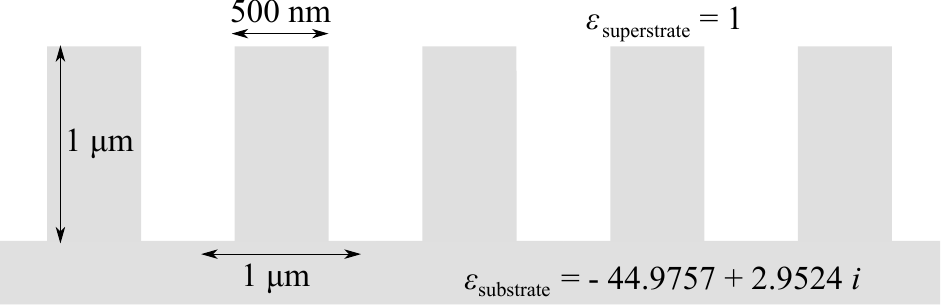}\\
  \caption{Rectangular groove grating: This pattern is repeatedly set up with a period $d$ = 1 $\mu m$. This grating has been studied by \cite{granet1999reformulation} and is one of our points of reference}
  \label{fig:Diffractive_Pattern_Geometry_rect}
  }
\end{figure}

We can refer to \cite{granet1999reformulation} in order to test the accuracy of
our method. The studied grating is isotropic, since we lack
numerical values in the literature in anisotropic cases. We compute
the following problem (cf.
Fig.~\ref{fig:Diffractive_Pattern_Geometry_rect}), as described in
\cite{Bao1} and \cite{granet1999reformulation}. The wavelength of the plane
wave is set to $1\,\mu m$ and is incoming with an angle of $\pi/6$
with respect to the normal to the grating.

\begin{table}[!ht]
\begin{center}
\begin{tabular}{|c|c|c|}
  \hline
  Maximum element size & $R_0^{\mathrm{TE}}$ & $R_0^{\mathrm{TM}}$ \\
 \hline
  $\lambda_0/(4\,\sqrt{\epsilon})$ & 0.7336765 & 0.8532342 \\
  $\lambda_0/(6\,\sqrt{\epsilon})$ & 0.7371302 & 0.8456592 \\
  $\lambda_0/(8\,\sqrt{\epsilon})$ & 0.7347466 & 0.8482817 \\
  $\lambda_0/(10\,\sqrt{\epsilon})$ & 0.7333739 & 0.850071 \\
  $\lambda_0/(12\,\sqrt{\epsilon})$ & 0.7346569 & 0.8494844 \\
  $\lambda_0/(14\,\sqrt{\epsilon})$ & 0.7341944 & 0.8483238 \\
  $\lambda_0/(16\,\sqrt{\epsilon})$ & 0.7342714 & 0.8484774 \\
  \hline
  Result given by \cite{granet1999reformulation} & 0.7342789    &   0.8484781 \\
  \hline
\end{tabular}
  \caption{Reflected efficiencies versus mesh refinement. Note
that the efficiencies are properly computed (two significant digits)
even for a rather coarse mesh.} \label{tab:R_0_biblio}
\end{center}
\end{table}

We present the $R_0$ efficiency (cf. Table \ref{tab:R_0_biblio}) in
both cases of polarization versus the mesh refinement. So
we have a good agreement to the reference values, and the accuracy
reached is independent from the polarization case.

\subsubsection{Experiment set up based on existing materials}\label{sec:aniso_example}
\begin{figure}[ht!]
\center{
  \includegraphics[width=0.7 \textwidth,draft=\logic]{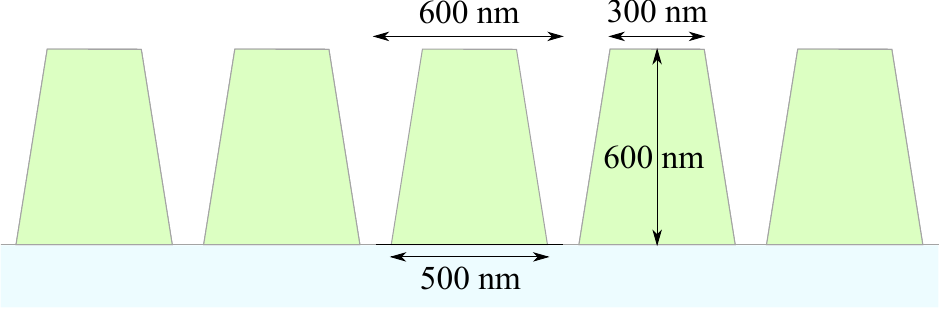}\\
  \caption{Diffractive element pattern. This element is made of aragonite for which the dielectric tensor is given
  by Eq.~(\ref{eq:aragonite_dielectric_tensor}) and is deposited on a silica substrate with a period $d=600\nm$.}
  \label{fig:Diffractive_Pattern_Geometry}
  }
\end{figure}

The method proposed in this section is adapted to $z$--anisotropic materials,
such as transparent $\textrm{CaCO}_3$ \cite{these_tayeb}, $\textrm{LiNbO}_3$
\cite{ohkawa} or Ni:YIG \cite{kono2004novel} and lossy 
CoPt or CoPd \cite{zhou1992dielectric}. Let us now consider a trapezoidal (cf. Fig.~\ref{fig:Diffractive_Pattern_Geometry}) anisotropic grating made of
aragonite ($\textrm{CaCO}_3$) deposited on an isotropic substratum
($\textrm{SiO}_2$, $\varepsilon_{\,\mathrm{SiO_2}}=2.25$). Along the
anisotropic crystal axis, its dielectric tensor can be written as
follows \cite{these_tayeb}:
\begin{equation}
\tenseps_{\,\textrm{CaCO}_3}=%
\left( %
\begin{array}{ccc}
  2.843 & 0 & 0 \\
  0 & 2.341 & 0 \\
  0 & 0 & 2.829
\end{array}
\right) %
\quad \hbox{and} \quad
\tensmu_{\,\mathrm{CaCO_3}}=%
\left( %
\begin{array}{ccc}
  \mu_{0} & 0 & 0 \\
  0 & \mu_{0} & 0 \\
  0 & 0 & \mu_{0}
\end{array}
\right) \;
\end{equation}

Now let's assume  that the natural axis of our aragonite grating are
rotated by $45^{\,\circ}$ around the grating infinite dimension. The
dielectric tensor becomes:
\begin{equation}
\tenseps_{\,\textrm{CaCO}_3}^{45^{\,\circ}}=%
\left( %
\begin{array}{ccc}
  2.592 & 0.251 & 0 \\
  0.251 & 2.592 & 0 \\
  0 & 0 & 2.829
\end{array}
\right) \label{eq:aragonite_dielectric_tensor}
\end{equation}
We shall here remind that our method remains strictly the same
whatever the diffractive element geometry is. The 2D computational
domain is bounded along the $y$--axis by the PMLs and along the $x$
since we consider only one pseudo period. We propose to calculate
the diffractive efficiencies at $\lambda_0 = 633\,\textrm{nm}$ in both
polarization cases TE and TM, and for different incoming incidences
($0^{\,\circ}$, $20^{\,\circ}$ and $40^{\,\circ}$). Since both
$\tensmu$ and $\tenseps$ are Hermitian, the whole incident energy is
diffracted and the sum of theses efficiencies ought to be equal to
the incident energy, which will stand for validation of our
numerical calculation.

Finally, the resulting bounded domain is meshed with a maximum mesh
element side size of $\lambda_0/10\,\sqrt{\varepsilon}$.
Efficiencies are still post-processed in accordance with the
calculation presented {section \ref{section:diff_eff2D}.}

\begin{figure}[!ht]
    \begin{center}
            {\large \textbf{TM}}  \qquad \qquad \qquad \qquad \qquad \qquad \qquad \qquad  {\large \textbf{TE}} \\
                {\includegraphics[width=0.4\textwidth,draft=\logic]{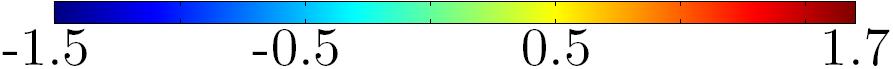}\label{subfig:TE_echelle}}
                \qquad
            {\includegraphics[width=0.4\textwidth]{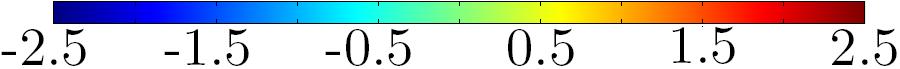}\label{subfig:TM_echelle}}\\
    \subfloat[$\RE\{e\}$ in $V/m$ at $\theta_0\,=\,0^{\,\circ}$]{\includegraphics[width=0.4 \textwidth,draft=\logic]{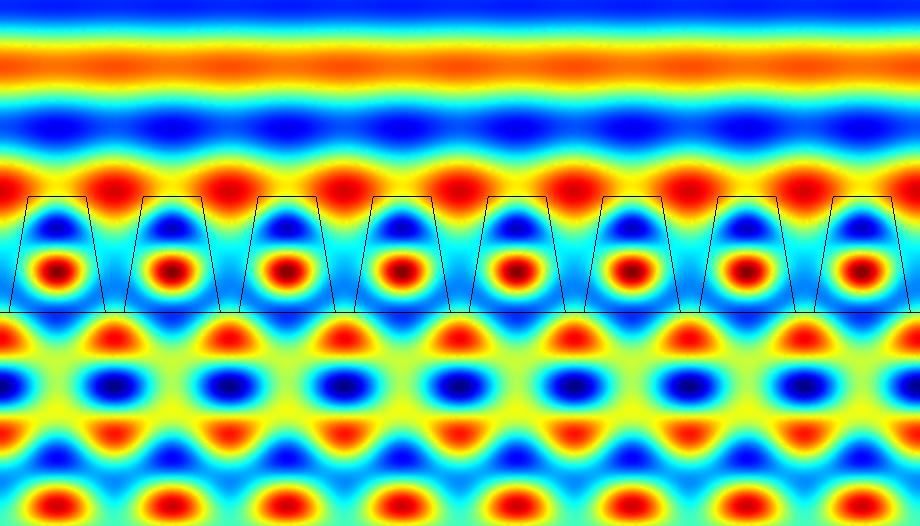}\label{subfig:TM_0}}
    \qquad
    \subfloat[$\RE\{h\}$ in $A/m$ at
    $\theta_0\,=\,0^{\,\circ}$]{\includegraphics[width=0.4
    \textwidth,draft=\logic]{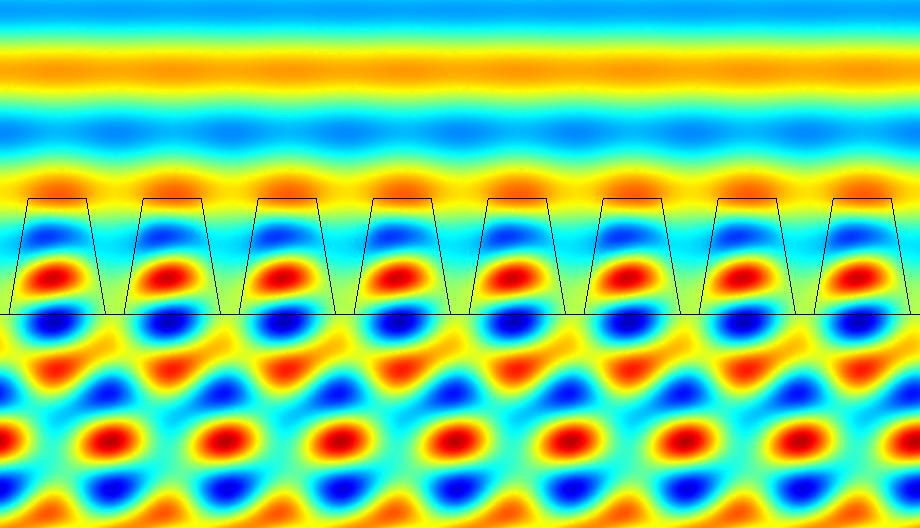}\label{subfig:TE_0}}\\
    \subfloat[$\RE\{e\}$ in $V/m$ at $\theta_0\,=\,20^{\,\circ}$]{\includegraphics[width=0.4 \textwidth,draft=\logic]{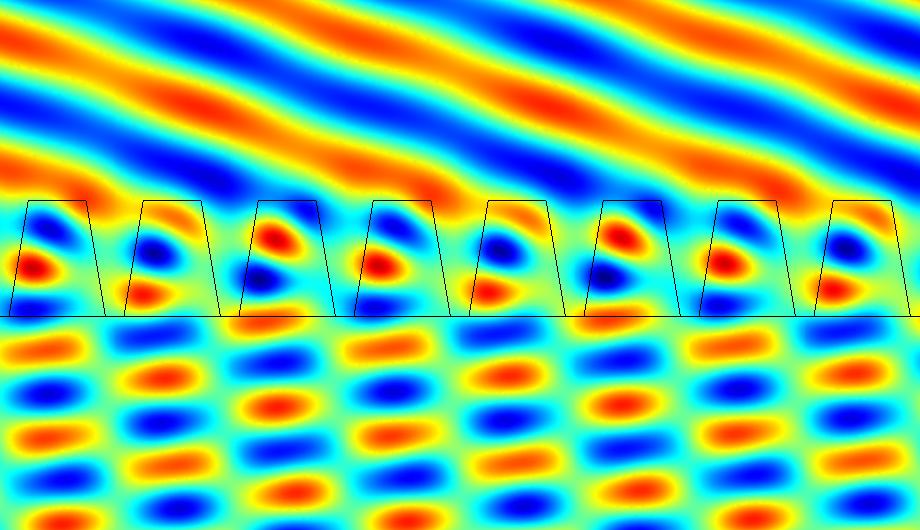}\label{subfig:TM_20}}
    \qquad
    \subfloat[$\RE\{h\}$ in $A/m$ at
    $\theta_0\,=\,20^{\,\circ}$]{\includegraphics[width=0.4
    \textwidth,draft=\logic]{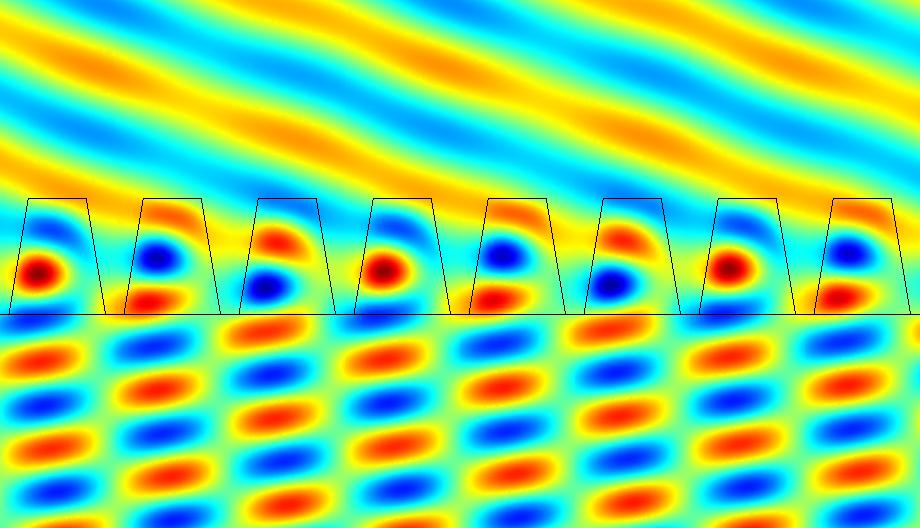}\label{subfig:TE_20}}\\
    \subfloat[$\RE\{e\}$ in $V/m$ at $\theta_0\,=\,40^{\,\circ}$]{\includegraphics[width=0.4 \textwidth,draft=\logic]{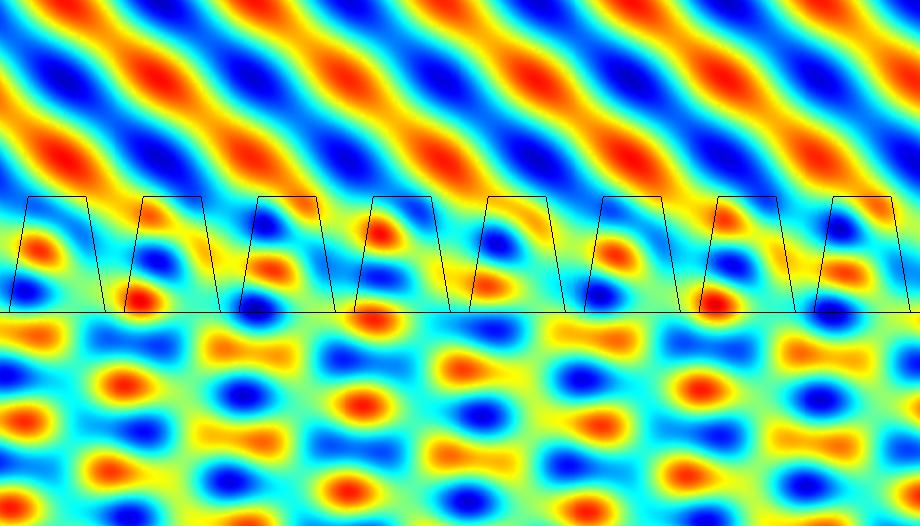}\label{subfig:TM_40}}
    \qquad
    \subfloat[$\RE\{h\}$ in $A/m$ at
    $\theta_0\,=\,40^{\,\circ}$]{\includegraphics[width=0.4
    \textwidth,draft=\logic]{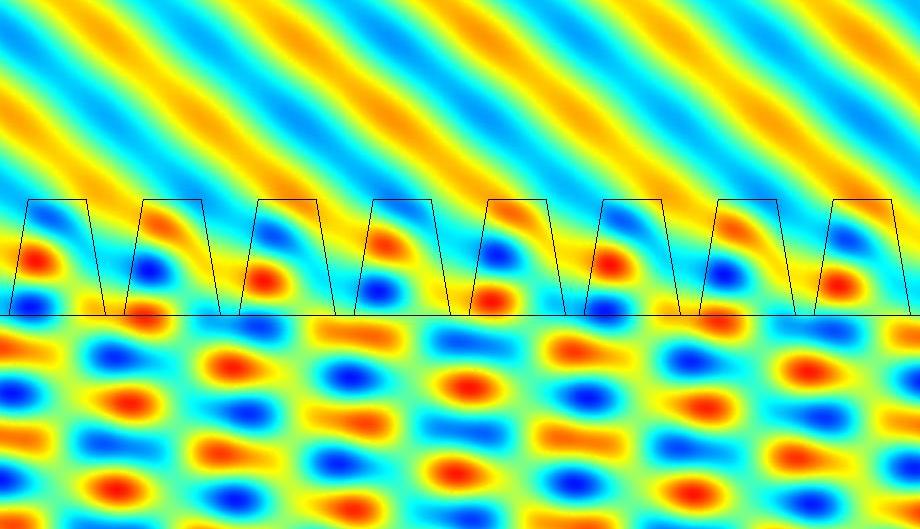}\label{subfig:TE_40}}\\
\caption{Real part of the total calculated field depending on
$\theta_0$ and the polarization case} \label{fig:cartes_re_E}
    \end{center}
\end{figure}
\begin{table}[!ht]
    \begin{center}
        \begin{tabular}{|c||c|c|c|c||c|c|c||c|}
            \hline
            \textbf{TM} & $T_{-2}$ & $T_{-1}$ & $T_{0}$ & $T_{1}$ & $R_{-1}$ & $R_{0}$ & $R_{1}$ & total\\
            \hline
            $0^{\,\circ}$ &- &  0.203133 &0.585235 &0.203138 &-  & 0.008473  &-  &0.999978\\
            \hline
            $20^{\,\circ}$& - &  0.399719 &0.575625 &0.004643  &0.004412 & 0.015630 & -  &1.000029\\
            \hline
            $40^{\,\circ}$&0.025047 & 0.420714 &0.493491 &-  & 0.002541 & 0.058238 & -  &1.000031\\
            \hline
            \hline
            \textbf{TE} & $T_{-2}$ & $T_{-1}$ & $T_{0}$ & $T_{1}$ & $R_{-1}$ & $R_{0}$ & $R_{1}$ & total\\
            \hline
            $0^{\,\circ}$ &- &  0.322510 &0.538165 &0.124722& -  & 0.014683 & -  &1.000080\\
            \hline
            $20^{\,\circ}$&- &  0.538727& 0.444403& 0.000369&  0.005372&  0.011180&  - &1.000051\\
            \hline
            $40^{\,\circ}$&0.012058 & 0.434191& 0.541090& - &  0.005032 & 0.007686 & - &1.000057\\
            \hline
        \end{tabular}\\
    \end{center}
\caption{Transmitted and reflected efficiencies of propagative
orders deduced from field maps shown Fig.~\ref{fig:cartes_re_E}}
\label{tab:anisotropic case}
\end{table}
At normal incidence, the $h$ field in the TE case (cf. Fig.~\ref{subfig:TE_0}) is non symmetric whereas the $e$ field in the TM
case is (cf. Fig.~\ref{subfig:TM_0}). This is illustrated by the
obvious non-symmetry of $T_{-1}^{\mathrm{TE}}$ and
$T_1^{\mathrm{TE}}$ (cf. Table \ref{tab:anisotropic case}: 0.322510
versus 0.124722!), whereas
$T_{-1}^{\mathrm{TM}}=T_{1}^{\mathrm{TM}}=0.20313$.
\newpage
\subsubsection{A non trivial geometry}\label{sec:marsu}
Since the beginning of this chapter, we have laid great stress upon 
the independence of the method towards the geometry of the pattern. 
But we have considered so far diffractive objects of simple trapezoidal section. 
Let us tackle a way more challenging case and see what this approach is made of.

\begin{figure}[!ht]
\centering
\includegraphics[width=.93\textwidth,draft=\logic]{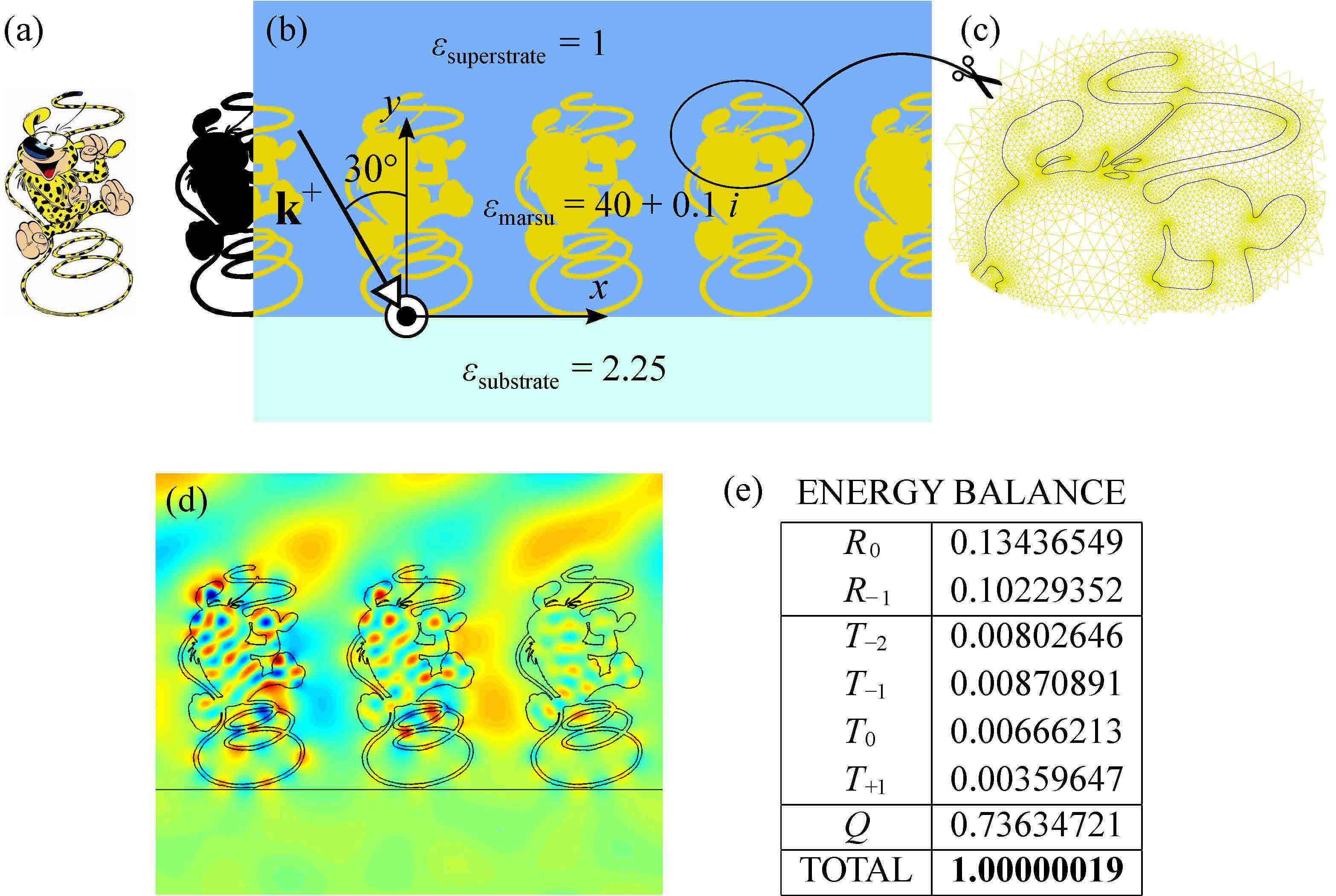}
\caption{(a) Initial contrasted image. (b) Proposed set up. (c) Sample mesh. (d) $\Re e\{E_z\}$ in $V/m$. (e) Energy balance of the problem. }
\label{fig:marsu}
\end{figure}

We can obtain an quite winding shape by extracting the contrast contour of an arbitrary image 
(see Fig.~\ref{fig:marsu}a-\ref{fig:marsu}b). The contour is approximated by a set of splines,
and the resulting domain is finely meshed (Fig.~\ref{fig:marsu}c). Finally, as shown in 
Fig.~\ref{fig:marsu}b, the formed pattern ($h^g/\lambda_0=1.68$), breathing in free space
($\varepsilon_\textrm{substrat}=1$), is supposed to be periodically repeated $d/\lambda_0=1.26$
on a plane ground of glass ($\varepsilon_{\textrm{SiO}_2}=2.25$). The element is considered to
be \enquote{made of} a lossy material of high optical index ($\varepsilon_{\textrm{marsu}}=40+0.1\,i$).
The response of this system to a incident \textit{s}-polarized plane wave at oblic incidence 
($\theta_0=-30^\circ$) is finally calculated. The real part of the quasi-periodic total field
is represented in Fig.~\ref{fig:marsu}d for three periods. 

Indeed, we do not have any tabulated data available to check our results. But what we do have is 
a pretty reliable consistency check through the computation of the energy balance described by 
Eqs.~(\ref{eq:diffefficiency}) and~(\ref{eq:losses2D}). As shown in Fig.~\ref{fig:marsu}e, we obtain 
at least 7 significative digits on the energetic values. The total balance of 1.00000019 is computed 
taking into account (i) values of the total field inside the diffractive elements, (ii) values of the 
diffracted field at altitudes spanning the whole (modeled) superstrate, (iii) values of the total field 
at altitudes spanning the entire (modeled) substrate. Finally, (iv) the calculated field $u_2^d$ also 
nicely decays exponentially inside both PML. These four points allow us to check  \textit{a posteriori} 
the validity of the field everywhere in the computation cell.

\toreadover
\newpage
\subsection{Dealing with Wood anomalies using Adaptative PML} \label{sec:adaptative_pml}
As we have noticed at the end of Section~\ref{subsec:PML}, PMLs based on 
\enquote{traditional coordinate stretching} are inefficient for
periodic problems when dealing with grazing angles of diffracted orders, \textit{i.e.} when the
frequency is near a Wood's anomaly (\cite{wood1902rcu,rayleigh1907nrc}), leading to spurious
reflexions and thus numerical pollution of the results. An important question in designing
absorbing layers is thus the choice of their parameters: The PML thickness and the absorption
coefficient. To this aim, adaptative formulations have already been set up, most of them employing a
posteriori error estimate \cite{Chen2005, Bao1, schadle2007domain}. In this section, we propose
Adaptative PMLs (APMLs) with a suitable coordinate stretching, depending both on incidence and
grating parameters, capable of efficiently absorbing propagating waves with nearly grazing
angles. This section is dedicated to the mathematical formulation used to
determine PML parameters adapted to any diffraction orders. We provide at the end a
numerical example of a dielectric slit grating showing the relevance of our approach in
comparison with classical PMLs. 

\subsubsection{Skin depth of the PML} \label{impl_pml}

\begin{figure}[!ht]
\centering
\includegraphics[width=0.5\columnwidth,draft=\logic]{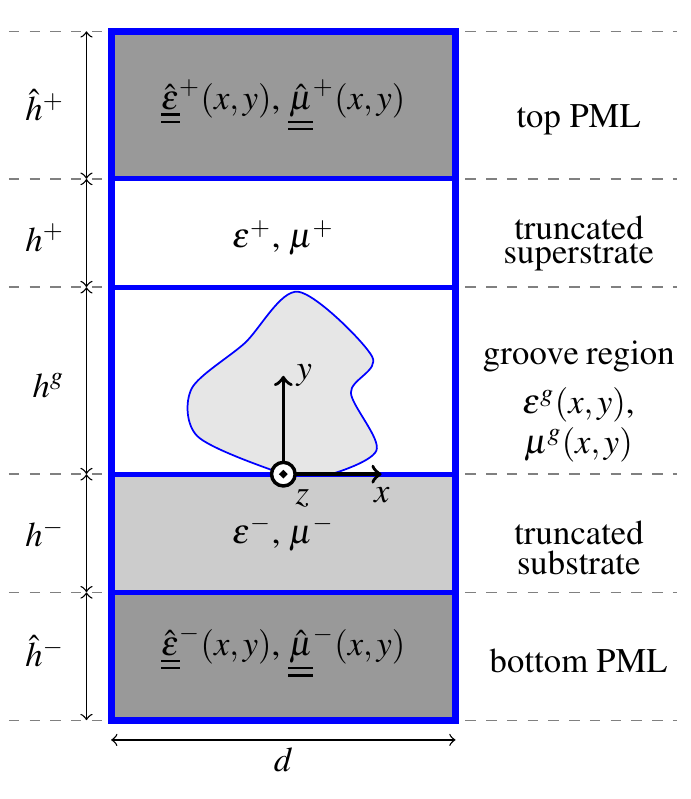}
\caption{The basic cell used for the FEM computation of the diffracted field $u_2^d$.}
\label{cell}
\end{figure}
\toreadover
As explained in Section~\ref{section:diff_eff2D}, the diffracted field $u^d$ can 
be expanded as a Rayleigh expansion, \textit{i.e.} into an infinite sum of propagating 
and evanescent plane waves called diffraction orders. As detailed at the end of 
Section~\ref{subsec:PML}, we are now in position to 
rewrite easily the expression of, say, a transmitted diffraction order into the substrate. 
Similar considerations also apply to the reflected orders in the top PML. Combining Eq.~(\ref{eq:wave_in_PML}) and~(\ref{eq:coefffourier2a}) lead 
to the expression $u_{n,s}^{-}(y_c)$ of a transmitted propagative order inside the PML:

%
%
%
%
%
%
%

\begin{equation*}
 u_{n,s}^{-}(y_c)=u_n^-(y(y_c))=t_n\e^{-\ic\beta_n^-[y^t+\zeta^-(y_c-y^t)]}.
\end{equation*}

The non oscillating part of this function is given by:
\begin{equation*}
U_n^{-}(y)=t_n\exp{\left((\beta'^{,-}_n\zeta''^{,-}+\beta''^{,-}_n\zeta'^{,-})y_c\right)},
\end{equation*}
where $\beta_n^-=\beta'^{,-}_n+\ic\beta''^{,-}_n$. For a propagating order we have 
$\beta'^{,-}_n>0$ and $\beta'^{,-}_n=0$, while for an evanescent order 
$\beta'^{,-}_n=0$ and $\beta''^{,-}_n>0$. It is thus sufficient to take 
$\zeta'^{,-}>0$ and $\zeta''^{,-}>0$ to ensure the exponential decay to 
zero of the field inside the PML \textit{if it was of infinite extent}.
But, of course, for practical purposes, the thickness of the PML is finite 
and has to be suitably chosen. Two pitfalls must be avoided:
\begin{enumerate}
 \item The PML thickness is chosen too small compared to the skin depth. 
 	As a consequence, the electromagnetic wave cannot
 	be considered as vanishing: An incident electromagnetic \enquote{sees the bottom of the PML}. 
	In other words, this PML of finite thickness is no longer reflection-less.
\item The PML thickness is chosen much larger than the skin depth.
	 In that case, a significant part of the PML is not useful,
 	 which gives rise to the resolution of linear systems of unnecessarily large dimensions.
\end{enumerate}
Then remains to derive the skin depth, $l_n^-$, associated with the propagating order $n$.
This characteristic length is defined as the depth below the PML at which the field falls to $1/e$ of its value near the surface:
 \begin{equation*}
  U_n^{-}(y-l_n^-)=\frac{U_n^{-}(y)}{e}.
 \end{equation*}
Finally, we find $l_n^-=(\beta'^{,-}_n\zeta''^{,-}+\beta''^{,-}_n\zeta'^{,-})^{-1}$ and we define $l^-$ as the largest value among the $l_n^-$:
 \begin{equation*}
  l^-=\underset{n\in\mathbb{Z}}{\mathrm{max}}\,l_n^-.
 \end{equation*}
The height of the bottom PML region is set to $\hat{h}^-=10l^-$.

\subsubsection{Weakness of the classical PML for grazing diffracted angles}
\label{example_pb_classical_PML}
\begin{figure}[!ht]
\centering
\includegraphics[width=0.8\columnwidth]{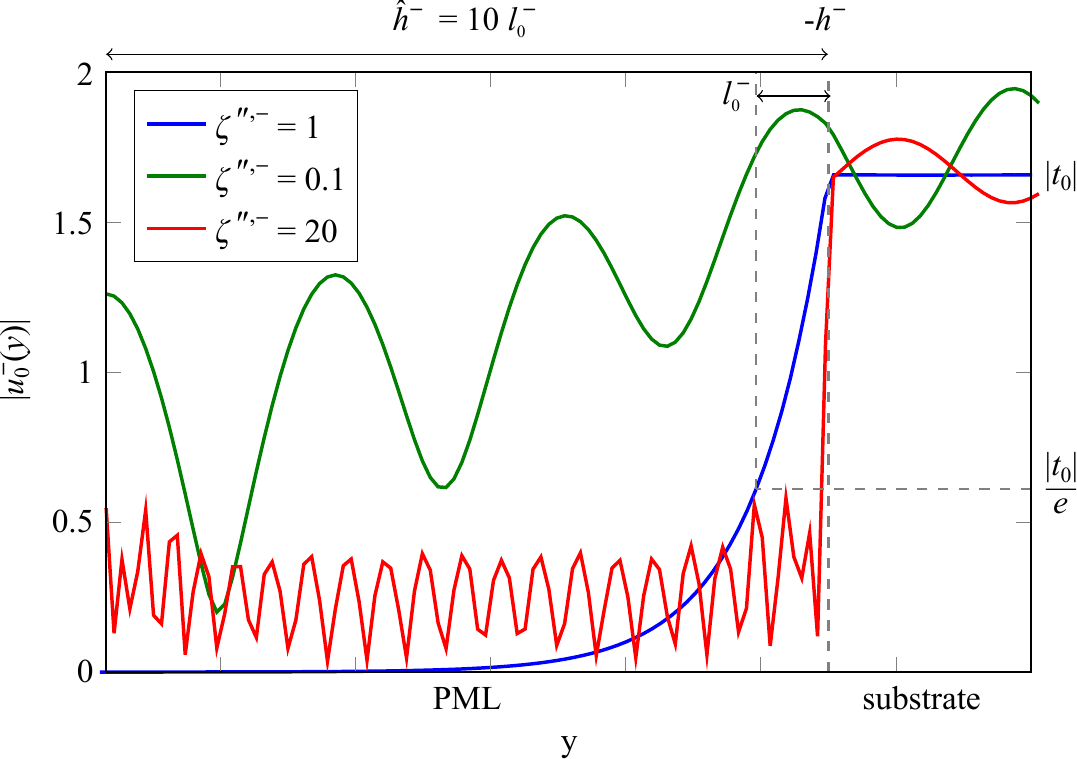}
\caption{Zero\textsuperscript{th} transmitted order by a grating with a rectangular cross section (see parameters in text, part \ref{example_pb_classical_PML}) for different values of
$\zeta''^{,-}$: blue line, $\zeta''^{,-}=1$, correct damping; green line, $\zeta''^{,-}=0.1$, underdamping; red line, $\zeta''^{,-}=20$, overdamping.}
\label{behaviour_classical_PML}
\end{figure}
Let us consider the (bottom) PML adapted to the substrate. Similar conclusions will hold for the top PML.
The efficiency of the classical PML fails for grazing diffracted angles, in other words when a given order appears/vanishes:
this is the so-called Wood's anomaly, well known in the grating theory. In mathematical terms, there exists $n_0$ such that
$\beta_{n_0}^-\simeq 0$. The skin depth of the PML then becomes very large. To compensate this, it is tempting to increase the
value of $\zeta''^{,-}$, but it would lead to spurious numerical reflections due to an overdamping. For a fixed value of $\hat{h}^-$,
if $\zeta''^{,-}$ is too weak, the absorption in the PMLs is insufficient and the wave is reflected on the outward boundary of the PML.
To illustrate these typical behaviors  (cf. Fig.~\ref{behaviour_classical_PML}), we compute the 
field diffracted by a grating with a rectangular cross section of height $h^g=1.5\micron$ and
width $L^g=3\micron$ with $\varepsilon^g=11.7$, deposited on a substrate with permittivity $\varepsilon^-=2.25$. 
The structure is illuminated by a \textit{p}-polarized plane wave of wavelength 
$\lambda_0=10\micron$ and of angle of incidence $\theta_0=10^\circ$ in the air ($\varepsilon^+=1$). All materials are 
non magnetic ($\mu_r=1$) and the periodicity of the grating is $d=4\micron$. We set $\hat{h}^-=10l_0^-$ and
$\zeta'^{,-}=1$.

\subsubsection{Construction of an adaptative PML}
\label{part_stretch}
To overcome the problems pointed out in the previous section, we propose a
coordinate stretching that rigorously treats the problem of Wood's anomalies.
The wavelengths \enquote{seen} by the system are very different depending on the order at stake:
\begin{itemize}
 \item if the diffracted angle $\theta_n$ is zero, the apparent wavelength $\lambda_0/\cos\theta_n$ is simply the incident wavelength,
 \item if the diffracted angle is near $\pm \pi/2$ (grazing angle), the apparent wavelength $\lambda_0/\cos\theta_n$ is very large.
\end{itemize}
Thus if a classical PML is adapted to one diffracted order, it will not be for another, and
vice versa. The idea behind the APML is to deal with each and every order when progressing in the absorbing medium.\\

Once again the development will be conducted only for the PML adapted to the substrate.
We consider a real-valued coordinate mapping $y_d(y)$, the final
complex-valued mapping is then $y_c(y)=\zeta^-y_d(y)$, with the complex constant
$\zeta^-$, with $\zeta'^{,-}>0$ and $\zeta''^{,-}>0$, accounting for the damping of the PML medium.

\begin{figure}[!ht]
\centering
\includegraphics[width=0.6\columnwidth]{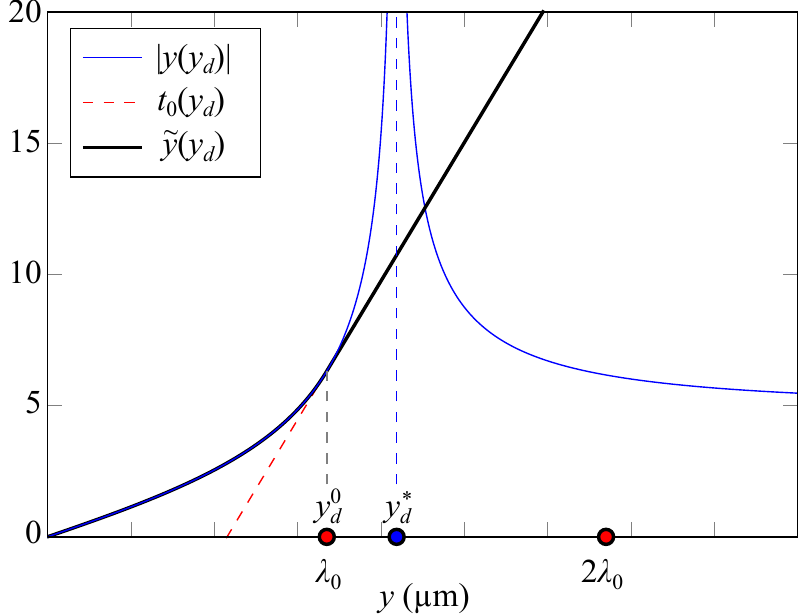}
\caption{Example of a coordinate mapping $\tilde{y}(y_d)$ used for the APML (black solid line). The graph of $y_d(y)$ (blue solid line) is continued
 by a straight line $t_0(y_d)$ tangent at $y_d^0$ (red dashed line) to avoid the singular behaviour at $y_d=y_{d}^\star$.}
\label{coef_stretch}
\end{figure}

We begin with transforming the equation ${\beta_n^\pm}^2={k^\pm}^2-{\alpha_n^\pm}^2$, so that
the function with integer argument $n\mapsto\beta_n^-$ becomes a function with real argument
continuously interpolated between the imposed integer values. Indeed, the geometric transformations associated to the PML has to
be continuous and differentiable in order to compute its Jacobian. To that extent, we choose the parametrization:
\begin{equation}
 \alpha(y_d)=\alpha_0+\frac{2\pi}{d}\frac{y_d}{\lambda_0},
\end{equation}
\noindent so that the application $\beta^-$ defined
by ${\beta^-(y_d)}^2=k_0^2\varepsilon^--{\alpha(y_d)}^2$ is continuous.
Thus, the propagation constant of the $n^{\text{th}}$ transmitted order is given by
$\beta_n^-=\beta^-(n\lambda_0)$. The key idea is to combine the complex stretching with
a real non uniform contraction (given by the continuous function $y(y_d)$, Eq.~(\ref{yyd})). 
This contraction is chosen in such a way that for each order $n$ there is a depth $y_d^n$ such 
that, around this depth, the apparent wavelength corresponding to the order in play is contracted
to a value close to $\lambda_0$. At that point of the PML, this order is perfectly absorbed thanks
to the complex stretch. We thus eliminate first the orders with quasi normal diffracted angles at
lowest depths up to grazing orders (near Wood's anomalies) which are absorbed at greater depths. 
In mathematical words, the translation of previous considerations on the real contraction can be expressed as:
\begin{equation}
\exp{[-\ic\beta^-(y_d)y(y_d)]}=\exp{(-\ic k_0 y_d)}
\end{equation}
The contraction $y(y_d)$ is thus given by:
\begin{equation}
 y(y_d)=\frac{k_0y_d}{\beta^-(y_d)}=\frac{y_d}{\sqrt{\varepsilon^--(\sin{\theta_0}+y_d/d)^2}}
\label{yyd}
\end{equation}
\noindent
The function $y(y_d)$ has two poles, denoted
$y_{d,\pm}^\star=d(\pm\sqrt{\varepsilon^-}-\sin{\theta_0})$. When $y_{d,\pm}^\star=\pm n\lambda_0$ with $n\in\mathbb{N}^\star$,
$\beta^-(y_{d,\pm}^\star)=\beta^-(\pm n\lambda_0)=\beta_{\pm}^-=0$, i.e.\ we are on a Wood's anomaly associated
with the appearance/disappearance of the $\pm n^{\text{th}}$ transmitted order. We now search for the nearest
point to $y_{d,\pm}^*$ associated
with a Wood's anomaly, denoting:
\begin{equation*}
\begin{cases}
 n_+^\star/ \quad D_+=\underset{n_+^\star\in\mathbb{N}^\star}{\mathrm{min}}\,|y_{d,+}^\star-n_+^\star\lambda_0|\\
\\
 n_-^\star/ \quad D_-=\underset{n_-^\star\in\mathbb{N}^\star}{\mathrm{min}}\,|y_{d,-}^\star+n_-^\star\lambda_0| 
\end{cases} \,\, .
\end{equation*}
In a second step, we look for the point $y_d^0=n^\star\lambda_0$ such that:
\begin{equation}
 n^\star/ \quad D=\underset{n^\star\in\{n_+^\star,n_-^\star \}}{\mathrm{min}}\,(D^+,D^-)\,\,.
\end{equation}
To avoid the singular behaviour at $y_d=y_{d,\pm}^\star$,
we continue the graph of the function $y_d(y)$ by a straight line tangent at $y_d^0$, which equation is $t_0(y_d)=s(y_d^0)(y_d-y_d^0)+y(y_d^0)$,
where $s(y_d)=\frac{\partial y}{\partial y_d}(y_d)$ is the so-called stretching coefficient.
The final change of coordinate is then given by~:
\begin{equation}
 \tilde{y}(y_d)=
\begin{cases}
y(y_d) \mbox{ for } y_d\leq y_d^0\\
\\
t_0(y_d) \mbox{ for } y_d>y_d^0.
\end{cases}
\end{equation}
Figure \ref{coef_stretch} shows an example of this coordinate mapping.
Eventually, the complex stretch $s_y$ used in Eq.~(\ref{tenspml}) is given by:
\begin{equation}
 s_y(y_d)=\zeta^-\frac{\partial \tilde{y}}{\partial y_d}(y_d).
\end{equation}

Equipped with this mathematical formulation, we can tailor a layer that is doubly
perfectly matched:
\begin{itemize}
 \item to a given medium, which is the aim of the PML technique, through Eq.~(\ref{equ_tranform}),
\item to all diffraction orders, through the stretching coefficient $s_y$, which
depends on the characteristics of the incident wave and on opto-geometric parameters
 of the grating.
\end{itemize}

\subsubsection{Numerical example}

We now apply the method described in the preceding parts to design an adapted bottom PML for the same example
 as in part \ref{example_pb_classical_PML}. The parameters are the same, and we choose the wavelength
of the incident plane wave close to the Wood's anomaly related to the $+1$ transmitted order
($\lambda_0=0.999 y_{d,+}^\star$).Moreover, we set the length of the PML $\hat{h}^-=1.1y_{d,+}^\star$ and choose absorption coefficients
$\zeta^{+}=\zeta^{-}=1+\ic$. For both cases (PML and APML), parameters are alike, the only difference being the complex stretch $s_y$.

\begin{figure}[!ht]
\centering
\includegraphics[width=0.8\columnwidth]{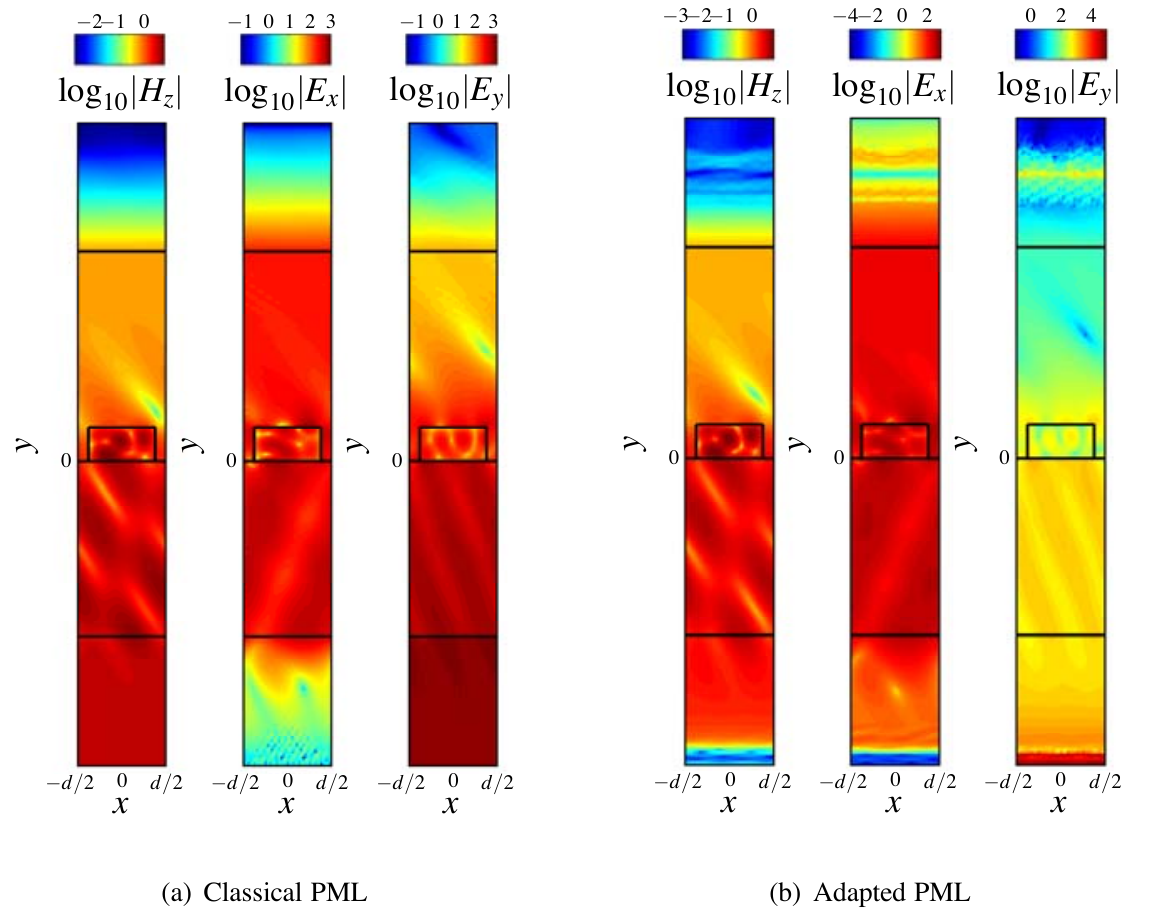}
\caption{Field maps of the logarithm of the norm of $H_z$, $E_x$ and $E_y$ for the dielectric slit grating at $\lambda_0=0.999 y_{d,+}^\star$ (same parameters as in part \ref{example_pb_classical_PML}). (a): classical PML with inefficient damping of $H_z$ in the bottom PML. (b): APML where the $H_z$ field is correctly damped in the bottom PML. For both cases the thickness of the PML is $\hat{h}^-=1.1y_{d,+}^\star$.}
\label{comp_PMLs}
\end{figure}

\begin{figure}[!ht]
\centering
\includegraphics[width=0.7\columnwidth]{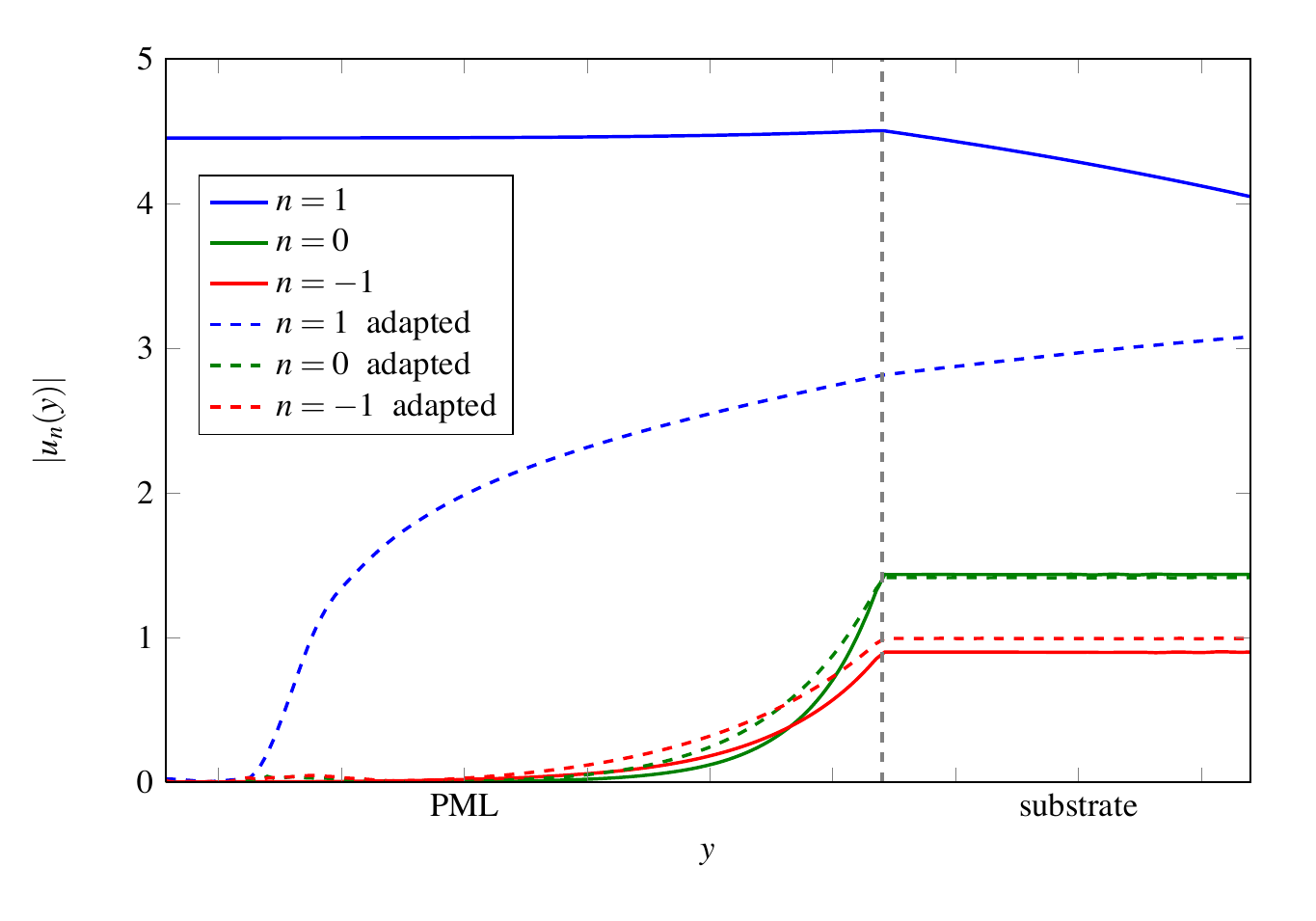}
\caption{Modulus of the $u_n$ for the three propagating orders with adapted (dashed lines) and classical PMLs (solid lines).
Note that the classical PMLs are efficient for all orders except for the grazing one ($n=1$) as expected. This drawback is bypassed when using the adaptative PML. }
\label{behaviour_adapted_PML}
\end{figure}

 The field maps of the norm of $H_z$, $E_x$ and $E_y$ are plotted in logarithmic scale on Fig.~\ref{comp_PMLs}, for the case of a classical PML  and our APML.
We can observe that the field $H_z$ that is effectively computed is clearly damped in the bottom APML (leftmost on Fig.~\ref{comp_PMLs}(b))
 whereas it is not in the standard case (leftmost on Fig.~\ref{comp_PMLs}(a)), causing spurious reflections on the outer boundary.
The fields $E_x$ and $E_y$ are deduced from $H_z$ thanks to Maxwell's equations. The high values
of $E_y$ at the tip of the APML (rightmost on Fig.~\ref{comp_PMLs}(b)) are due to very high values of the optical
equivalent properties of the APML medium (due to high values of $s_y$), which does not affect the accuracy of the computed field within the domain of interest.\\
Another feature of our approach is that it efficiently absorbs the grazing diffraction order, as illustrated on Fig.~\ref{behaviour_adapted_PML}: the $+1$ transmitted order does not decrease in the standard
PML (blue solid line), and reaches a high value at $y=-\hat{h}^-$, whereas the same order
tends to zero as $y\rightarrow-\hat{h}^-$ in the case of the adapted PML (blue dashed line).\\
To further validate the accuracy of the method, we compare the diffraction efficiencies computed by our FEM formulation with PML and APML to those obtained by another method. We choose the Rigorous Coupled Wave Analysis (RCWA), also known as the Fourier Modal Method (FMM, \cite{li3}). For the chosen parameters, only the $0$\textsuperscript{th} order is propagative in reflexion and the orders $-1$, $0$ and $+1$ are non evanescent in transmission.
We can also check the energy balance $B=R_0+T_{-1}+T_{0}+T_{+1}$ since there is no lossy medium in our example. Results are reported in Table \ref{comp_PML_RCWA}, and show a good agreement of the FEM with APML with the results from RCWA. On the contrary, if classical PML are used, the diffraction efficiencies are less accurate compared to those computed with RCWA. Checking the energy balance leads the same conclusions: the numerical result is perturbed by the reflection of the waves at the end of the PML if it is not adapted to the situation of nearly grazing diffracted orders.\\

\begin{table}[!ht]
 \centering
\begin{tabular}{|c|c|c|c|c|c|}
  \hline
   & $R_0$ & $T_{-1}$ & $T_{0}$ & $T_{+1}$ & $B$ \\
  \hline
  RCWA & 0.1570 & 0.3966 &  0.1783 & 0.2680 & 0.9999 \\
  FEM + APML & 0.1561 & 0.3959 & 0.1776 & 0.2703 & 0.9999 \\
  FEM + PML & 0.1904 & 0.4118 & 0.1927 & 0.2481 & 1.0430 \\
  \hline
\end{tabular}
\caption{Diffraction efficiencies $R_0$, $T_{-1}$, $T_{0}$ and $T_{+1}$ of the four propagating orders, and energy balance $B=R_0+T_{-1}+T_{0}+T_{+1}$, computed by three methods: RCWA (line 1), FEM formulation with APML (line 2), FEM formulation with classical PML (line 3).}
\label{comp_PML_RCWA}
\end{table}

\begin{figure}[!ht]
\centering
\includegraphics[width=0.6\columnwidth]{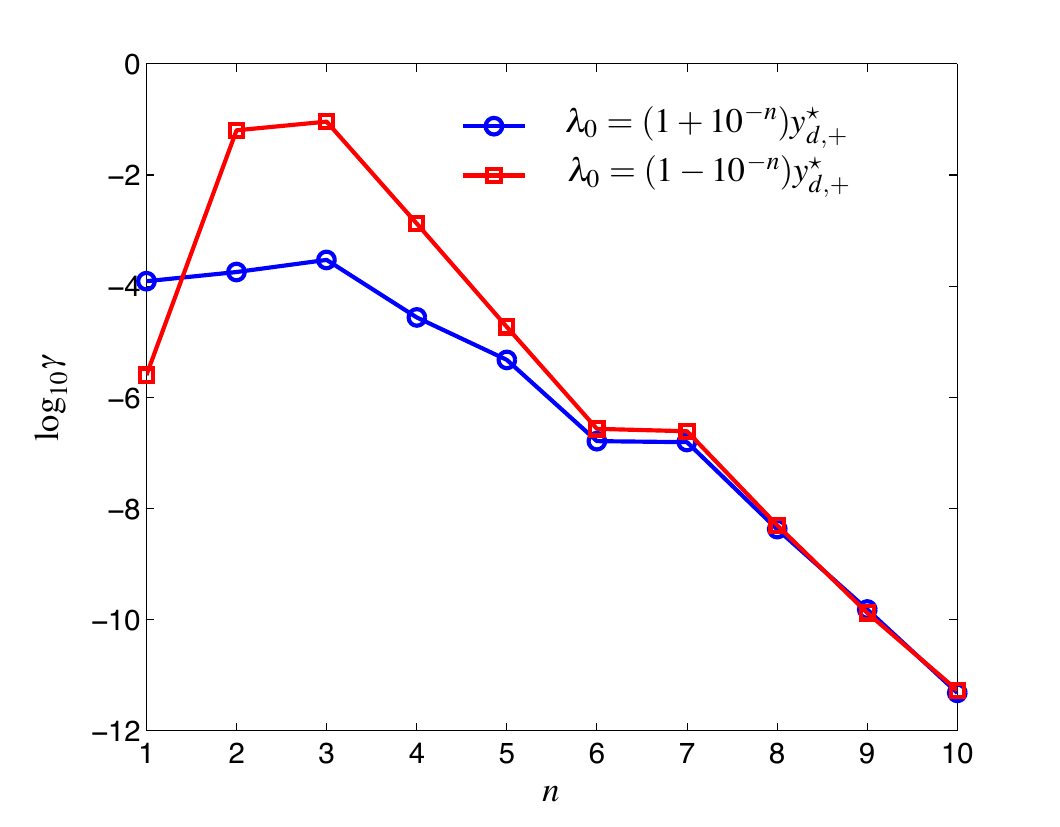}
\caption{Mean value of the norm of $H_z$
along the outer boundary of the bottom PML $\gamma=\langle|H_z(-\hat{h}^-)|\rangle_x$, for $\lambda_0$ approaching the
Wood's anomaly $y_{d,+}^\star$ by inferior values ($\lambda_0=(1-10^{-n})y_{d,+}^\star$, red squares)
 and by superior value ($\lambda_0=(1+10^{-n})y_{d,+}^\star$, blue circles) as a function of $n$.}
\label{comp_PML_lambda}
\end{figure}

Eventually, to illustrate the behavior of the adaptative PML when the incident wavelength gets
 closer to a given Wood's anomaly, we computed the mean value of the norm of $H_z$
along the outer boundary of the bottom PML $\gamma=\langle|H_z(-\hat{h}^-)|\rangle_x$, when $\lambda_0=(1+10^{-n})y_{d,+}^\star$
and $\lambda_0=(1-10^{-n})y_{d,+}^\star$, for $n=1,2,...10$. The results are shown in
Fig.~\ref{comp_PML_lambda}. As the wavelength gets closer to $y_{d,+}^\star$,
$\gamma$ first increases but for $n>3$, it decreases exponentially. However, in all cases, the value
of $\gamma$ remains small enough to ensure the efficiency of the PMLs.\\

\subsection{Concluding remarks}
A novel FEM formulation was adapted to the analysis of
z-anisotropic gratings relying on a rigorous treatment of the plane
wave sources problem through an equivalent radiation problem with
localized sources. The developed approach presents the advantage of
being very general in the sense that it is applicable to every
conceivable grating geometry.

Numerical experiments based on existing materials at normal and
oblique incidences in both TE and TM cases showed the efficiency and
the accuracy of our method. We demonstrated we could generate strongly
imbalanced symmetric propagative orders in the TE polarization case
and at normal incidence with an aragonite grating on a silica
substratum.

We also introduced the adaptative PML for grazing incidences configurations.
It based on a complex-valued coordinate stretching that deals with
grazing diffracted orders, yielding an efficient absorption of the field inside the PML.
We provided an example in the TM polarization case (but similar results hold for the TE case),
illustrating the efficiency of our method. The value of the magnetic field on the
outward boundary of the PML remains small enough to consider there is no spurious reflection.
The formulation is used with the FEM but can be applied to others numerical methods. Moreover, the generalization to the vectorial
three-dimensional case is straightforward: the recipes given in this last section do work irrespective of the dimension and whether
the problem is vectorial.

In the next section, the scalar formulation adapted to mono-dimensional gratings is extended 
to the the most general case of 
bi-dimensional grating embedded in an arbitrary multilayered dielectric stack with arbitrary incidence.

\bla


\newpage
\section{Diffraction by arbitrary crossed-gratings : a vector Finite Element formulation}
\label{sec:3D}

\subsection{Introduction}
\todo 

In this section, we extend the method detailed in Sec.~\ref{sec:2D} 
to the most general case of vector diffraction by an arbitrary crossed gratings.
The main advantage of the Finite Element Method lies in its native ability to handle 
unstructured meshes, resulting in a build-in accurate discretization of oblique edges. 
Consequently, our approach remains independent of the shape of the diffractive element, 
whereas other methods require heavy adjustments depending on whether
the geometry of the groove region presents oblique edges
(\textit{e.g.} RCWA \cite{popov2002staircase}, FDTD\dots). 
In this section, for the sake of clarity, we recall again the rigorous procedure allowing 
to deal with the issue of the plane wave sources through an equivalence of the
diffraction problem with a radiation one whose sources are localized
inside the diffractive element itself, as already proposed in Sec.~\ref{sec:2D} \cite{Demesy1,demesy:058002}.

This approach combined with the use of second order edge elements
allowed us to retrieve with a good accuracy the few numerical
academic examples found in the literature. Furthermore, we provide a
new reference case combining major difficulties such as a non
trivial toroidal geometry together with strong losses and a high
permittivity contrast. Finally, we discuss computation time and
convergence \textcolor{Map2}{as a} function of the mesh refinement
as well as the choice of the direct solver.
\bla

\subsection{Theoretical developments}
\subsubsection{Set up of the problem and notations}\label{part:3D_def_notations}
We denote by $\x$, $\y$ and $\z$ the unit vectors of the axes of an
orthogonal coordinate system $Oxyz$. We only deal with time-harmonic
fields; consequently, electric and magnetic fields are represented
by the complex vector fields $\bE$ and $\bH$, with a time dependance
in $\exp(-i\, \omega\, t)$. Note that incident light is now propagating 
along the $z$-axis, whereas $y$-axis was used in the 2D case. 

\begin{figure}[!ht]\centering
  \includegraphics[width=7cm,draft=\logic]{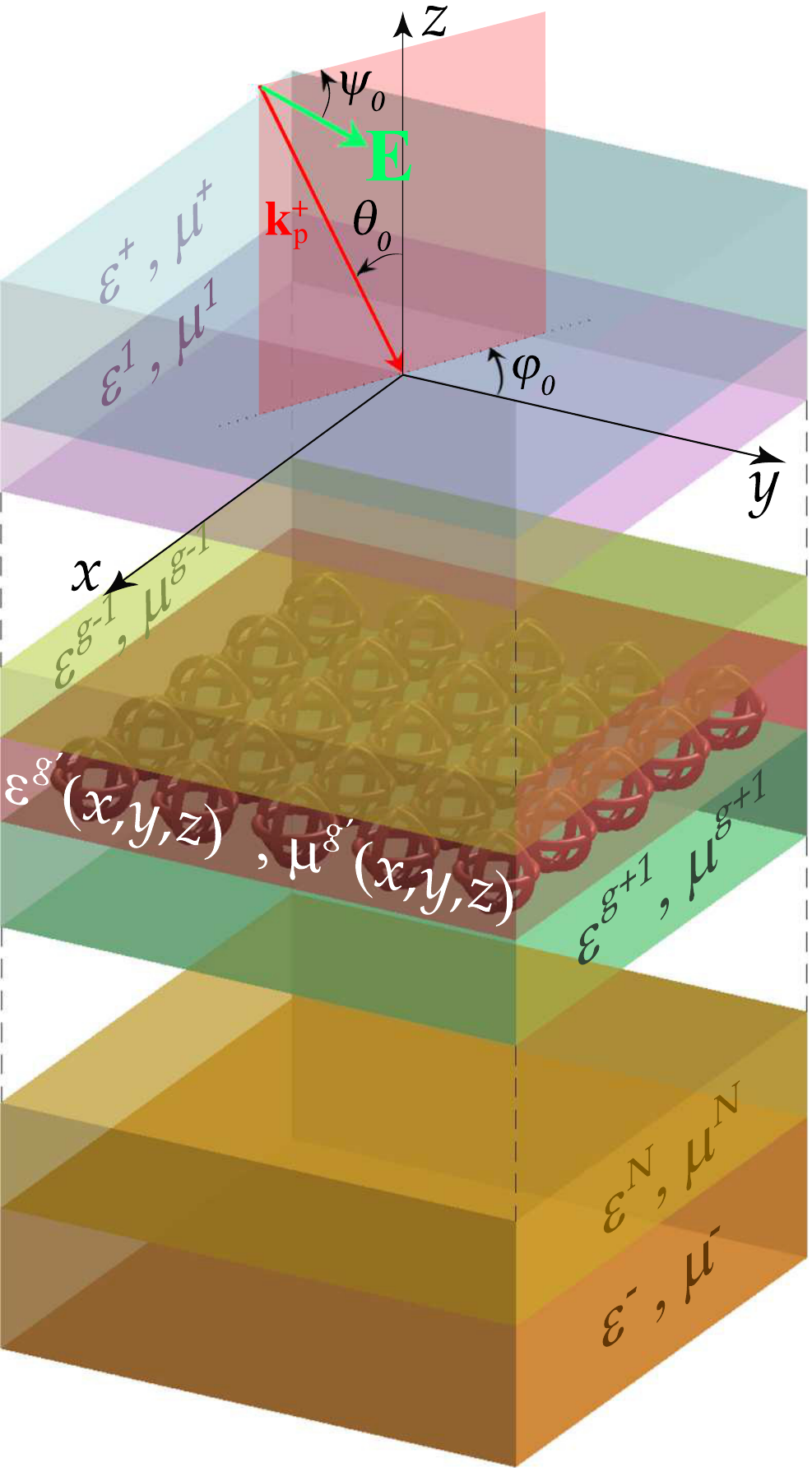}\caption{Scheme and notations of the studied bi-gratings.} \label{fig:schema_3D_multi}
\end{figure}

Besides, in this section, for the sake of simplicity, the materials
are assumed to be isotropic and therefore are optically
characterized by their relative permittivity $\varepsilon$ and
relative permeability $\mu$ \textcolor{Clarte}{(note that the
inverse of relative permeabilities are denoted here
$\nucol$)}. It is of importance to note that lossy materials can be
studied, the relative permittivity and relative permeability being
represented by complex valued functions. The crossed-gratings we are dealing with 
can be split into the following
regions as suggested in Fig.~\ref{fig:schema_3D_multi}:
\begin{itemize}
\item%
    \textit{The  superstrate} ($z> z_0$) is
    supposed to be homogeneous, isotropic and lossless, and therefore characterized
    by its relative permittivity $\varepsilon^+$ and its relative
    permeability
    $\mu^+(=1/\nucol^+)$ and we denote $k^+:=k_0\, \sqrt{\varepsilon^+ \mu^+}$, where $k_0:=\omega/c$,%
\item%
    \textit{The multilayered stack} (\cor$z_N<z<z_0$\bla) is made of $N$ layers which are supposed to be
    homogeneous and isotropic, and therefore characterized by their relative permittivity
    $\varepsilon^n$, their relative permeability
    $\mu^n(=1/\nucol^n)$ and their thickness $e_n$. We denote
    $k_n:=~k_0\,\sqrt{\varepsilon^n\,\mu^n}$ for $n$ integer between $1$ and $N$.
\item
    \textit{The groove region} (\cor$z_{g}<z<z_{g-1}$\bla), which is embedded in
    the layer indexed $g$ $(\varepsilon^g,\mu^g)$ of the previously
    described domain, is heterogeneous. Moreover
    the method does work irrespective of whether the
    diffractive elements are homogeneous: The permittivity and permeability can
    vary continuously (gradient index gratings) or discontinuously (step
    index gratings). This region is thus characterized by the
    scalar fields $\varepsilon^{g'}(x,y,z)$ and $\mu^{g'}(x,y,z)(=1/\nucol^{g'}(x,y,z))$. The
    groove periodicity along the $x$--axis, respectively (resp.) $y$--axis, is denoted
    $d_x$, resp. $d_y$, in the sequel.
\item%
    \textit{The substrate} ($z<z_N$) is supposed to be homogeneous
    and isotropic and therefore characterized by its relative
    permittivity $\varepsilon^-$ and its relative permeability
    $\mu^-(=1/\nucol^-)$ and we denote $k^-:=k_0\, \sqrt{\varepsilon^- \mu^-}$,%
\end{itemize}
Let us emphasize the fact that the method principles remain
unchanged in the case of several diffractive patterns made of
distinct geometry and/or material.

The incident field on this structure is denoted:
\begin{equation}\label{eq:def_Einc}
    \bE^{\mathrm{inc}} =\textbf{A}_0^e\;\textrm{exp}(i \,\textbf{k}^+_p\cdot\textbf{r})
\end{equation}
with
\begin{equation}\label{eq:def_kplus}
    \textbf{k}^+ = \left[ \begin{array}{l} \alpha_0\\\beta_0\\\gamma_0
    \end{array} \right] = k^+\,\left[ \begin{array}{l} -\sin \theta_0 \,\cos \varphi_0 \\-\sin \theta_0 \,\sin \varphi_0 \\-\cos \theta_0
    \end{array} \right]
\end{equation}
and
\begin{equation}\label{eq:def_Azero}
    \textbf{A}_0^e = \left[ \begin{array}{l} E^0_x\\E^0_y\\E^0_z
    \end{array} \right] =A^e\,\left[ \begin{array}{l} \cos \psi_0 \, \cos \theta_0  \,\cos \varphi_0  - \sin \psi_0 \, \sin \varphi_0 \\\cos \psi_0 \, \cos \theta_0 \, \sin \varphi_0  + \sin \psi_0 \, \cos \varphi_0 \\-\cos \psi_0 \, \sin \theta_0
    \end{array} \right] ,
\end{equation}
where $\varphi_0\in[0,2\pi]$, $\theta_0\in[0,\pi/2]$ and
$\psi_0\in[0,\pi]$ (polarization angle).

We recall here the diffraction problem: finding the solution of Maxwell equations in harmonic
regime \textit{i.e.} the unique solution ($\bE,\bH$) of:
\begin{subequations}\label{eq:Maxwell2}
\begin{numcases}{}
\curl\, \bE=i\, \omega\, \mu_0\, \mu \, \bH \label{eq:MaxwellrotE2}\\
\curl\, \bH=-i\, \omega\, \varepsilon_0\, \varepsilon \, \bE
\label{eq:MaxwellrotH2}
\end{numcases}
\end{subequations}
such that the diffracted field satisfies the so-called
\textit{Outgoing Waves Condition} (OWC \cite{zolla1996method} ) and
where $\bE$ and $\bH$ are quasi-bi-periodic functions with respect
to $x$ and $y$ coordinates.

One can choose to calculate arbitrarily $\bE$, since $\bH$ can be
deduced from Eq.~(\ref{eq:MaxwellrotE2}). The diffraction problem
amounts to looking for the unique solution $\bE$ of the so-called
vectorial Helmholtz propagation equation, deduced from
Eqs.~(\ref{eq:MaxwellrotE2},\ref{eq:MaxwellrotH2}):
\begin{equation}\label{eq:defLvect}
    \Lvect_{\varepsilon,\nucol}:=-\rot\left(\nucol\,\rot\bE\right) + k_0^2\,\varepsilon\,\bE =
    \textbf{0}
\end{equation}
such that the diffracted field satisfies an OWC and where $\bE$ is a
quasi-bi-periodic function with respect to $x$ and $y$ coordinates.
\subsubsection{From a diffraction problem to a radiative one with localized
sources}\label{part:radiation_pb} According to
Fig.~\ref{fig:schema_3D_multi}, the scalar relative permittivity
$\varepsilon$ and \textcolor{Clarte}{inverse} permeability $\nucol$
fields associated to the studied diffractive structure can be
written using complex-valued functions defined by part and taking
into account the notations adopted in
Sec.~\ref{part:3D_def_notations}:
\begin{equation}\label{eq:def_eps_global}
\upsilon(x,y,z):= \left \{
\begin{array}{lcccc}
  \upsilon^+ & \hbox{for} & z>z_0 &&\\
  \upsilon^n & \hbox{for} & z_{n-1}>z>z_n & \hbox{with} & 1\leq n < g\\
  \upsilon^{g'}(x,y,z) & \hbox{for} & z_{g-1}>z>z_g &&\\
  \upsilon^n & \hbox{for} & z_{n-1}>z>z_n& \hbox{with} & g<n \leq N\\
  \upsilon^- & \hbox{for} & z<z_N&&
\end{array}
\right .
\end{equation}
\noindent with $\upsilon=\{ \varepsilon, \nucol \}\quad , \quad
z_0=0 \quad \textrm{and} \quad z_n = -\sum_{l=1}^{n}e_l \quad
\textrm{for} \quad 1\leq n \leq N$.

\noindent It is now convenient to introduce two functions defined by
part $\varepsilon_1$ and $\nucol_1$ corresponding to the associated
multilayered case (\textit{i.e.} the same stack without any
diffractive element) constant over $Ox$ and $Oy$:
\begin{equation}\label{eq:def_eps1}
\upsilon_1(x,y,z):= \left \{
\begin{array}{lcccc}
  \upsilon^+ & \hbox{for} & z>0\\
  \upsilon^n & \hbox{for} & z_{n-1}>z>z_n & \hbox{with} & 1\leq n \leq N\\
  \upsilon^- & \hbox{for} & z<z_N
\end{array}
\right .
\end{equation}
\noindent with $\upsilon=\{ \varepsilon, \nucol \}$.

\noindent We denote by $\bE_0$ the restriction of
$\bE^{\mathrm{inc}}$ to the superstrate region:
\begin{equation}
\bE_0:= \left \{
\begin{array}{ccc}
  \bE^{\mathrm{inc}} & \hbox{for} & z>z_0 \\
  \mathbf{0} & \hbox{for} & z\leq z_0
\end{array}
\right .%
\end{equation}
\noindent We are now in a position to define more explicitly the
vector diffraction problem that we are dealing with in this
section. It amounts to looking for the unique vector field $\bE$
solution of:
\begin{equation}\label{eq:defLOWC_vect}
    \Lvect_{\varepsilon,\nucol}(\bE)=\mathbf{0} \quad \hbox{such that $\bE^d:=\bE-\bE_0$
satisfies an OWC.}
\end{equation}
\noindent In order to reduce this diffraction problem to a radiation
one, an intermediary vector field denoted $\bE_1$ is necessary and
is defined as the unique solution of:
\begin{equation}\label{eq:defL1_vect}
    \Lvect_{\varepsilon_1,\nucol_1}(\bE_1)=\mathbf{0} \quad \hbox{such that $\bE_1^d:=\bE_1-\bE_0$ satisfies an OWC.}
\end{equation}
\noindent The vector field $\bE_1$ corresponds to an \textit{ancillary
problem} associated to the \textit{general vectorial case of a
multilayered stack} which can be calculated \textit{independently}.
This general calculation is seldom treated in the literature, we
present a development in Appendix. Thus $\bE_1$ is from now on
\textit{considered as a known} vector field. It is now apropos to
introduce the unknown vector field $\bE_2^d$, simply defined as the
difference between $\bE$ and $\bE_1$, which can finally be
calculated thanks to the FEM and:
\begin{equation}\label{eq:E2d}
\bE_2^d:= \bE - \bE_1 =\bE^d-\bE_1^d\; .
\end{equation}
It is of importance to note that the presence of the superscript $d$
is not fortuitous: As a difference between two diffracted fields
(Eq.~(\ref{eq:E2d}), $\bE_2^d$ satisfies an OWC which is of prime
importance in our formulation. By taking into account these new
definitions, Eq.~(\ref{eq:defLOWC_vect}) can be written:
\begin{equation}\label{eq:radiating_vect}
    \Lvect_{\varepsilon,\nucol}(\bE_2^d) = -\Lvect_{\varepsilon,\nucol}(\bE_1)
\; ,
\end{equation}
\noindent where the right-hand member is a vector field which can be
interpreted as \textit{a known vectorial source term
$-\Svect_1(x,y,z)$ whose support is localized inside the diffractive
element itself}. To prove it, let us introduce the null term defined
in Eq.~(\ref{eq:defL1_vect}) and make the use of the linearity of
$\Lvect$, which leads to:
\begin{equation}
    \Svect_1:=\Lvect_{\varepsilon,
    \nucol}(\bE_1)=\Lvect_{\varepsilon,
    \nucol}(\bE_1)-\underbrace{\Lvect_{\varepsilon_1,
    \nucol_1}(\bE_1)}_{=\mathbf{0}}=\Lvect_{\varepsilon-\varepsilon_1, \nucol -
    \nucol_1}(\bE_1) \; .
\end{equation}
\subsubsection{Quasi-periodicity and weak formulation}\label{part:form_faible_3D}
The weak form is obtained by multiplying scalarly
Eq.~(\ref{eq:defLOWC_vect}) by weighted vectors $\bE'$ chosen among
the ensemble of quasi-bi-periodic vector fields of $L^2(\rot)$
(denoted $L^2\left(\rot,(d_x,d_y),\bk\right)$) in $\Omega$:
\begin{equation}\label{eq:R_vect_def}
    \Rvect_{\varepsilon,\nucol}(\bE,\bE')=\int_\Omega -\rot\left(\nucol\,\rot\bE\right)\cdot\overline{\bE'} + k_0^2\,\varepsilon\,\bE\cdot\overline{\bE'}\,\mathrm{d}\Omega
\end{equation}
\noindent Integrating by part Eq.~(\ref{eq:R_vect_def}) and making
the use of the Green-Ostrogradsky theorem lead to:
\begin{equation}\label{eq:weak}
    \Rvect_{\varepsilon,\nucol}(\bE,\bE')=\int_\Omega -\nucol\,\rot\bE\cdot\rot\overline{\bE'} +
    k_0^2\,\varepsilon\,\bE\cdot\overline{\bE'}\,\mathrm{d}\Omega -
    \int_{\partial \Omega}\left(\bn \times (\nucol\,\rot\bE)\right)\cdot\overline{\bE'}\,\mathrm{d}S
\end{equation}
where $\bn$ refers to the exterior unit vector normal to the surface
$\partial \Omega$ enclosing $\Omega$.

The first term of this sum concerns the volume behavior of the
unknown vector field whereas the right-hand term can be used to set
boundary conditions (Dirichlet, Neumann or so called quasi-periodic
Bloch-Floquet conditions).

The solution $\bE_2^d$ of the \textit{weak form associated to the
diffraction problem}, expressed in its previously defined
\textit{equivalent radiative form} at Eq.~(\ref{eq:radiating_vect}),
is the element of $L^2\left(\rot,(d_x,d_y),\bk\right)$ such that:

\begin{equation}\label{eq:sol_vect}
\forall \bE' \in L^2(\curl,d_x,d_y,\bk),
\,\Rvect_{\varepsilon,\nu}(\bE_2^d,\bE')=-\Rvect_{\varepsilon-\varepsilon_1,\nu-\nu_1}(\bE_1,\bE')
\, .
\end{equation}

In order to rigorously truncate the computation a set of Bloch
boundary conditions are imposed on the pair of planes defined by $(y
= -d_y/2,y = d_y/2)$ and $(x = -d_x/2,x = d_x/2)$. One can refer to
\cite{nicolet2004modelling} for a detailed implementation of Bloch
conditions adapted to the FEM. A set of Perfectly Matched Layers are
used in order to truncate the substrate and the superstrate along
$z$ axis (see \cite{OuldAgha1} for practical implementation of PML
adapted to the FEM). Since the proposed unknown $\bE_2^d$ is
quasi-bi-periodic and satisfies an OWC, this set of boundary
conditions is perfectly \cor reasonable\bla: $\bE_2^d$ is radiated
from the diffractive element towards the infinite regions of the
problem and decays exponentially inside the PMLs along $z$ axis. The
total field associated to the diffraction problem $\bE$ is deduced
at once from Eq.~(\ref{eq:E2d}).
\subsubsection{Edge or Whitney 1-form second order elements}
In the vectorial case, edge elements (or Whitney
forms) make a much more relevant choice \cite{dular1995discrete}
than nodal elements. Note that a lot of work (see for instance
\cite{ingelstrom2006new}) has been done on higher order edge
elements since their introduction by Bossavit
\cite{bossavit1989ees}. These elements are suitable to the
representation of vector fields such as $\bE_2^d$, by letting their
normal component be discontinuous and imposing the continuity of
their tangential components. Instead of linking the Degrees Of
Freedom (DOF) of the final algebraic system to the nodes of the
mesh, the DOF associated to edges (resp. faces) elements are the
\textit{circulations} (resp. \textit{flux}) of the unknown vector field
along (resp. across) its \textit{edges} (resp. \textit{faces}).

Let us consider the computation cell $\Omega$ together with its
exterior boundary $\partial \Omega$. This volume is sampled in a
finite number of tetrahedron according to the following rules: Two
distinct tetrahedrons have to either share a node, an edge or a face
or have no contact. Let us denote by $\mathcal{T}$ the set of
tetrahedrons, $\mathcal{F}$ the set of faces, $\mathcal{E}$ the set
of edges and $\mathcal{N}$ the set of nodes. In the sequel, one will
refers to the node $n=\{i\}$, the edge $e=\{i,j\}$, the face
$f=\{i,j,k\}$ and the tetrahedron $t=\{i,j,k,l\}$.

\begin{figure}[!ht]
    \begin{center}
        \includegraphics[width=6 cm,draft=\logic]{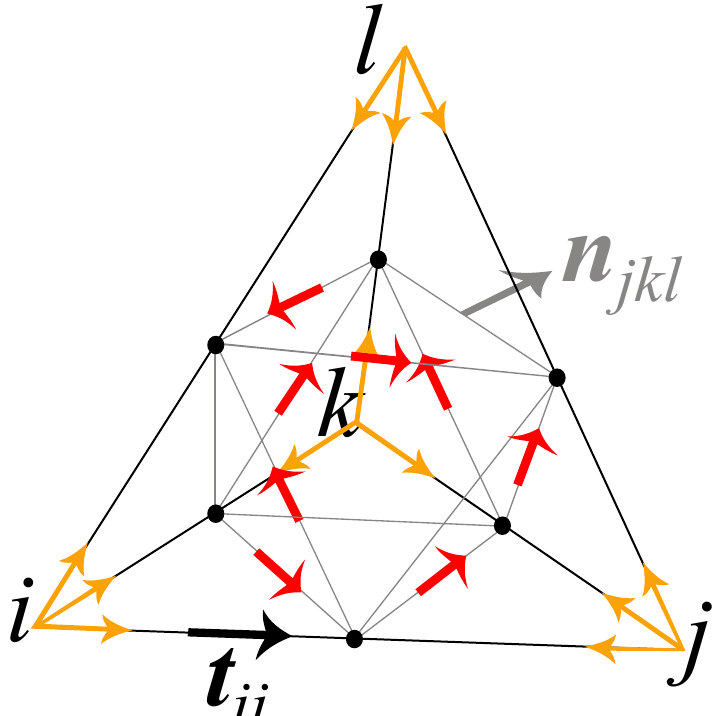}\caption{Degrees of freedom of a second order tetrahedral element.}\label{fig:EFordre2}
    \end{center}
\end{figure}
Twelve DOF (two for each of the six edges of a tetrahedron) are
classically derived from line integral of weighted projection of the
field $\bE_2^d$ on each oriented edge $e=\{i,j\}$:
\begin{equation}
            \left\{
                \begin{array}{c}
                    \vartheta_{ij} = \displaystyle \int_{i}^{j} \bE_2^d\cdot\bt_{ij}\,\lambda_i\,\textrm{d}l\\
                    \vartheta_{ji} = \displaystyle \int_{j}^{i} \bE_2^d\cdot\bt_{ji}\,\lambda_j\,\textrm{d}l
                \end{array}
            \right. ,
\end{equation}
where $\bt_{ij}$ is the unit vector and $\lambda_i$, the barycentric
coordinate of node $i$, is the chosen weight function.

According to Yioultsis \etal~\cite{yioultsis_magic_element}, a
judicious choice for the remaining DOF is to make the use of a
tangential projection of the 1-form $\bE_2^d$ on the face
$f=\{i,j,k\}$.
\begin{equation}
        \left\{
            \begin{array}{c}
                \vartheta_{ijk} =\displaystyle \int_f \left(\bE_2^d\times\bn_{ijk}^+\right)\cdot\grad\lambda_j\,\textrm{d}s\\
                \vartheta_{ikj} =\displaystyle \int_f \left(\bE_2^d\times\bn_{ijk}^-\right)\cdot\grad\lambda_k\,\textrm{d}s
            \end{array}
        \right. .
\end{equation}
The expressions for the shape functions, or basis vectors, of the
second order 1-form Whitney element are given by:
\begin{equation}
        \left\{
            \begin{array}{cl}
                \bw_{ij} &= (8\,\lambda_i^2-4\,\lambda_i)\,\grad \lambda_j + (-8\,\lambda_i\,\lambda_j+2\,\lambda_j)\,\grad \lambda_i\\
                \bw_{ijk}&= 16\,\lambda_i\,\lambda_j\,\grad \lambda_k
                -8\,\lambda_j\,\lambda_k\,\grad \lambda_i -
                8\,\lambda_k\,\lambda_i\,\grad \lambda_j
            \end{array}
        \right. .
\end{equation}
This choice of shape function ensures
\cite{yioultsis1996multiparametric} the following fundamental
property: every degree of freedom associated with a shape function
should be zero for any other shape function. Finally, an
approximation of the unknown $\bE_2^d$ projected on the shape
functions of the mesh $m$ ($\bE_2^{d,m}$) can be derived:
\begin{equation}\label{eq:decomp_champ_3D}
    \bE_2^{d,m} = \sum_{e\in\mathcal{E}} \vartheta_e\,\bw_e + \sum_{f\in\mathcal{F}}
    \vartheta_f\,\bw_f .
\end{equation}
Weight functions $\bE'$ (c.f. Eq.~(\ref{eq:sol_vect}) are chosen in
the same space than the unknown $\bE_2^d$,
$L^2(\rot,(d_x,d_y),\bk)$. According to the Galerkin formulation,
this choice is made so that their restriction to one bi-period
belongs to the set of shape functions mentioned above. Inserting the
decomposition of $\bE_2^d$ of Eq.~(\ref{eq:decomp_champ_3D}) in
Eq.~(\ref{eq:sol_vect}) leads to the final algebraic system which is
solved, in the following numerical examples, thanks to direct
solvers.
\subsection{Energetic considerations: Diffraction efficiencies and losses}
Contrarily to modal methods based on the determination of Rayleigh
coefficients, the rough results of the FEM are three complex
components of the vector field $\bE^d$ interpolated over the mesh of
the computation cell. Diffraction efficiencies are deduced from this
field maps as follows.

As a difference between two quasi-periodic vector fields (see
Eq.~(\ref{eq:defLOWC_vect})), $\bE^d$ is quasi-bi-periodic and its
components can be expanded as a double Rayleigh sum:
\begin{equation}\label{eq:sum_rayleigh}
    E^d_x (x,y,z) = \sum_{(n,m)\in \mathbb{Z}^2}
    u_{n,m}^{d,x}(z)\,e^{\textcolor{Map2}{i\,(\alpha_n\,x+\beta_m\,y)}} ,
\end{equation}
with $\alpha_n = \alpha_0 + \frac{2\,\pi}{d_x}\,n$, $\beta_m =
\beta_0 + \frac{2\,\pi}{d_y}\,m$ and
\begin{equation}\label{eq:u_n_3D}
    u_{n,m}^{d,x}(z) =
    \frac{1}{d_x\,d_y}\int_{-d_x/2}^{d_x/2}\int_{-d_y/2}^{d_y/2}E_x^d(x,y,z)\,e^{-i\,(\alpha_n\,x+\beta_m\,y)}\,\mathrm{d}x\,\mathrm{d}y\,.
\end{equation}
By inserting the decomposition of Eq.~(\ref{eq:sum_rayleigh}), which
is satisfied by $E_x^d$ everywhere but in the groove region, into
the Helmholtz propagation equation, one can express Rayleigh
coefficients in the substrate and the superstrate as follows:
\begin{equation}\label{eq:gamma_pm}
    u_{n,m}^{d,x}(z)= e_{n,m}^{x,p}\,e^{\,-i\,\gamma_{n,m}^+\,z} + e_{n,m}^{x,c} \,e^{\,i\,\gamma_{n,m}^+\,z}
\end{equation}
with $\gamma_{n,m}^{\pm^2} = k^{\pm^2} - \alpha_n^2 - \beta_m^2$,
where $\gamma_{n,m}$ (or $-i\,\gamma_{n,m}$) is positive. The
quantity $u_{n,m}^{d,x}$ is the sum of a propagative plane wave
(which propagates towards decreasing values of $z$, superscript $p$)
and of a counterpropagative one (superscript $c$). The OWC verified
by $\bE^d$ imposes:
\begin{equation}\label{eq:gamma_pm2}
\forall (n,m)\in \mathbb{Z}^2 \left \{
    \begin{array}{c}
        e_{n,m}^{x,p} = 0\quad \hbox{for}\quad z>z_0\\
        e_{n,m}^{x,c} = 0\quad \hbox{for}\quad z<z_N\\
    \end{array}
\right .
\end{equation}
Eq.~(\ref{eq:u_n_3D}) allows to evaluate numerically $e_{n,m}^{x,c}$
(resp. $e_{n,m}^{x,p}$) by double trapezoidal integration of a slice
of the complex component $E^d_x$ at an altitude $z_c$ fixed in the
superstrate (resp. substrate). It is well known that the mere
trapezoidal integration method is very efficient for smooth and
periodic functions (integration on one period). The same holds for
$E_y^d$ and $E_z^d$ components as well as their coefficients
$e_{n,m}^{y,\{c,p\}}$ and $e_{n,m}^{z,\{c,p\}}$.

\textcolor{Clarte}{The dimensionless expression of the efficiency of
each reflected and transmitted $(n,m)$ order \cite{noponen1} is
deduced from Eqs.~(\ref{eq:gamma_pm},\ref{eq:gamma_pm2}):}
\begin{equation}\label{eq:expression_vect_efficacity}
\left \{
    \begin{array}{ccc}
        R_{n,m}=\textcolor{Clarte}{\frac{1}{|A_e|^2}}\,\displaystyle \frac{\gamma_{n,m}^+}{\gamma_0}\,\be_{n,m}^c(z_c)\cdot\overline{\be_{n,m}^c(z_c)} \quad \hbox{for} \quad z_c>z_0\\
        T_{n,m}=\textcolor{Clarte}{\frac{1}{A_e^2}}\,\displaystyle \frac{\gamma_{n,m}^-}{\gamma_0}\,\be_{n,m}^p(z_c)\cdot\overline{\be_{n,m}^p(z_c)} \quad \hbox{for} \quad z_c<z_N\\
    \end{array}
\right . ,
\end{equation}
with $\be_{n,m}^{\{c,p\}}=
e_{n,m}^{x,\{c,p\}}\,\textbf{x}+e_{n,m}^{y,\{c,p\}}\,\textbf{y}+e_{n,m}^{z,\{c,p\}}\,\textbf{z}$.

\textcolor{Clarte}{Furthermore, normalized losses Q can be obtained
through the computation of the following ratio:
\begin{equation}\label{eq:losses}
    Q = \frac{\displaystyle \int_V \frac{1}{2}\,\omega\,\varepsilon_0\,\IM(\varepsilon^{g'})\,\bE\cdot\overline{\bE}\,\mathrm{d}V}
             {\displaystyle \int_S \frac{1}{2}\RE\{\bE_0\times\overline{\bH_0}\}\cdot \textbf{n}\,\mathrm{d}S}
    \,.
\end{equation}
The numerator in Eq.~(\ref{eq:losses}) clarifies losses in watts by
bi-period of the considered crossed-grating and are computed by
integrating the Joule effect losses density over the volume $V$ of
the lossy element. The denominator normalizes these losses to the
incident power, \textit{i.e.} the time-averaged incident Poynting
vector flux across one bi-period (a rectangular surface $S$ of area
$d_x\,d_y$ in the superstrate parallel to $Oxy$, whose normal
oriented along decreasing values of $z$ is denoted $\textbf{n}$).
Since $\bE_0$ is nothing but the plane wave defined at
Eqs.~(\ref{eq:def_kplus},\ref{eq:def_Azero}), this last term is
equal to
$(A_e^2\,\sqrt{\varepsilon_0/\mu_0}\,d_x\,d_y)/(2\,\textrm{cos}(\theta_0))$.}
Volumes and normal to surfaces being explicitly defined, normalized
losses losses $Q$ are quickly computed once $\bE$ determined and
interpolated between mesh nodes.

Finally, the accuracy and self-consistency of the whole calculation
can be evaluated by summing the real part of
transmitted and reflected efficiencies $(n,m)$ to normalized losses:
\begin{equation*}
Q+\sum_{(n,m)\in\mathbb{Z}^2}\RE\{R_{n,m}\}+\sum_{(n,m)\in\mathbb{Z}^2}\RE\{T_{n,m}\}\,
,
\end{equation*}
quantity to be compared to 1. The sole diffraction orders taken into
account in this conservation criterium correspond to propagative
orders whose efficiencies have a non-null real part. Indeed,
diffraction efficiencies of evanescent orders, corresponding to pure
imaginary values of $\gamma_{n,m}^\pm$ for higher values of $(n,m)$
(see Eq.~(\ref{eq:gamma_pm})) are also pure imaginary values as it
appears clearly in Eq.~(\ref{eq:expression_vect_efficacity}).
Numerical illustrations of such global energy balances are presented
in the next section.
\newpage
\subsection{Accuracy and convergence}
\subsubsection{Classical crossed gratings}
\begin{figure}[!ht]\centering
  \includegraphics[width=.9\textwidth,draft=\logic]{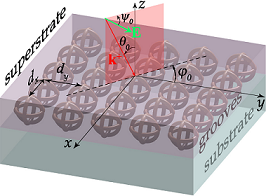}\caption{Configuration of the studied cases.} \label{fig:schema_3D}
\end{figure}

\textcolor{Map2}{There are only a few references in the literature
containing numerical examples.} For each of them, the problem only
consists of three regions (superstrate, grooves and substrate) as
summed up on Figure~\ref{fig:schema_3D}.
For the four selected cases, among six found in the literature,
published results are compared to ones given by our formulation of
the FEM. Moreover, in each case, a satisfying global energy balance
is detailed. Finally a new validation case combining all the
difficulties encountered when modeling crossed-gratings is proposed:
A non-trivial geometry for the diffractive pattern (a torus), made
of an arbitrary lossy material leading to a large step of index and
illuminated by a plane wave with an oblique incidence. Convergence
of the FEM calculation as well as computation time will be discussed
in Sec.~\ref{part:converg_3D}.

\begin{figure}[!ht]
    \begin{center}
        \subfloat[] {\includegraphics[width=.45\textwidth,draft=\logic]{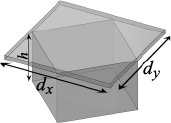}\label{fig:schema_checker}}
            \quad
        \subfloat[] {\includegraphics[width=.45\textwidth,draft=\logic]{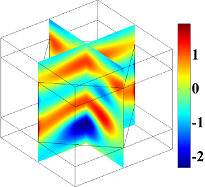}\label{fig:carte_checker}}
    \setlength{\captionmargin}{\mycaptionmargin}
    \caption{Diffractive element with vertical edges (a). $\RE \{E_x\}$ in V/m (b).}\label{fig:checker}
    \end{center}
\end{figure}

\newpage
\paragraph{Checkerboard grating}\label{part:damier}
In this example worked out by L.~Li \cite{li3}, the diffractive
element is a rectangular parallelepiped as shown
Fig.~\ref{fig:schema_checker} and the grating parameter highlighted 
in Fig.~\ref{fig:schema_3D} are the following:
$\varphi_0=\theta_0=0^{\,\circ}$, $\psi_0=45^{\,\circ}$, $d_x=d_y
=5\,\lambda_0\,\sqrt{2}/4$, $h=\lambda_0$,
$\varepsilon^+=\varepsilon^{g'}=2.25$ and $\varepsilon^- =
\varepsilon^{g}=1$.
\begin{table}[!ht]
\centering
    \begin{tabular}{lcc} \\ \hline
        &FMM \cite{li3} & FEM \\ \hline
        $T_{-1,-1}$         &0.04308&0.04333\\
        $T_{-1,0}$          &0.12860&0.12845\\
        $T_{-1,+1}$         &0.06196&0.06176\\
        $T_{0,-1}$          &0.12860&0.12838\\
        $T_{0,0}$           &0.17486&0.17577\\
        $T_{0,+1}$          &0.12860&0.12839\\
        $T_{+1,-1}$         &0.06196&0.06177\\
        $T_{+1,0}$          &0.12860&0.12843\\
        $T_{+1,+1}$         &0.04308&0.04332\\ \hline
        $\textcolor{Clarte}{\displaystyle \sum_{(n,m)\in\mathbb{Z}}} \RE\{R_{n,m}\}$ &    -  &0.10040\\ \hline
        $\textrm{TOTAL}$             &    -  &1.00000\\ \hline
    \end{tabular}\caption{Energy balance \cite{li3}.}\label{tab:result_checker}
\end{table}
\setlength{\captionmargin}{\mycaptionmargin} 

Our formulation of the FEM shows
good agreement with the Fourier Modal Method developed by L.~Li
(\cite{li3}, 1997) since the maximal relative difference between the
array of values presented in Table~\ref{tab:result_checker} remains
lower than $\textrm{10}^{\textrm{-3}}$. Moreover, the sum of the
efficiencies of propagative orders given by the FEM is very close to
1 in spite of the addition of all errors of determination upon the
efficiencies.

\newpage
\paragraph{Pyramidal crossed-grating}\label{part:pyramide}
In this example firstly worked out by Derrick \textit{et al.}
\cite{derrick1979cgt}, the diffractive element is a pyramid with
rectangular basis as shown Fig.~\ref{fig:schema_pyram} and the
grating parameters highlighted in Fig.~\ref{fig:schema_3D} are the
following: $\lambda_0 = 1.533$, $\varphi_0=45^{\,\circ}$,
$\theta_0=30^{\,\circ}$, $\psi_0=0^{\,\circ}$, $d_x=1.5$, $d_y=1$,
$h=0.25$, $\varepsilon^+ =\varepsilon^{g}= 1$ and
$\varepsilon^-=\varepsilon^{g'}=2.25$.
\setlength{\captionmargin}{\mycaptionmargin}
\begin{figure}[!ht]
    \begin{center}
        \subfloat[] {\includegraphics[width=.46\textwidth,draft=\logic]{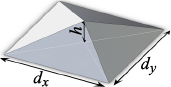}\label{fig:schema_pyram}}
            \quad
        \subfloat[] {\includegraphics[width=.46\textwidth,draft=\logic]{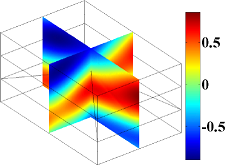}\label{fig:carte_pyram}}
    \setlength{\captionmargin}{\mycaptionmargin}
    \caption{Diffractive element with oblique edges (a). $\RE \{E_y\}$ in V/m (b).}\label{fig:pyram}
    \end{center}
\end{figure}
\begin{table}[!ht]
\centering
    \begin{tabular}{lcccccccc} \\ \hline
Given in &  \cite{derrick1979cgt}&  \cite{greffet1}&  \cite{brauer1}&  \cite{granet2}&FEM\\ \hline
            $R_{-1,0}$  & 0.00254 & 0.00207 & 0.00246 & 0.00249 & 0.00251\\
            $R_{0,0}$   & 0.01984 & 0.01928 & 0.01951 & 0.01963 & 0.01938\\
            $T_{-1,-1}$ & 0.00092 & 0.00081 & 0.00086 & 0.00086 & 0.00087\\
            $T_{0,-1}$  & 0.00704 & 0.00767 & 0.00679 & 0.00677 & 0.00692\\
            $T_{-1,0}$  & 0.00303 & 0.00370 & 0.00294 & 0.00294 & 0.00299\\
            $T_{0,0}$   & 0.96219 & 0.96316 & 0.96472 & 0.96448 & 0.96447\\
            $T_{1,0}$   & 0.00299 & 0.00332 & 0.00280 & 0.00282 & 0.00290\\ \hline
            $\textrm{TOTAL}$& 0.99855 & 1.00001 & 1.00008 & 0.99999 & 1.00004\\\hline
    \end{tabular}\caption{Comparison with the results given in \cite{derrick1979cgt,greffet1,brauer1,granet2}.}\label{tab:result_pyram}
\end{table}
Results given by the FEM show good agreement with ones of the C
method \cite{derrick1979cgt,granet2}, the Rayleigh method
\cite{greffet1} and the RCWA \cite{brauer1}. Note that, in this
case, some edges of the diffractive element are oblique.
\newpage
\paragraph{Bi-sinusoidal grating}\label{part:sinus}
In this example worked out by Bruno \textit{et al.} \cite{bruno1}, the
surface of the grating is bi-sinusoidal (see
Fig.~\ref{fig:schema_sinus}) and described by the function $f$
defined by:
\begin{equation}\label{eq:surf_bisinus}
    f(x,y) = \frac{h}{4}\left[\cos\left(\frac{2\,\pi\,x}{d}\right)+\cos\left(\frac{2\,\pi\,y}{d}\right)\right]
\end{equation}
The grating parameters \etal highlighted in Fig.~\ref{fig:schema_3D} are
the following: $\lambda_0 = 0.83$,
$\varphi_0=\theta_0=\psi_0=0^{\,\circ}$, $d_x=d_y=1$, $h=0.2$,
$\varepsilon^+ = \varepsilon^g=1$ and
$\varepsilon^-=\varepsilon^{g'}=4$.
\begin{figure}[!ht]
    \begin{center}
        \subfloat[] {\includegraphics[width=.45\textwidth,draft=\logic]{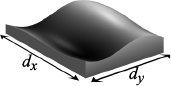}\label{fig:schema_sinus}}
            \quad
        \subfloat[] {\includegraphics[width=.45\textwidth,draft=\logic]{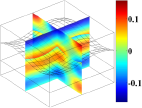}\label{fig:carte_sinus}}
    \setlength{\captionmargin}{\mycaptionmargin}
    \caption{Diffractive element with oblique edges (a). $\RE \{E_z\}$ in V/m (b).}\label{fig:sinus}
    \end{center}
\end{figure}
\begin{table}[!ht]
\centering
    \begin{tabular}{lcc} \hline
            &\cite{bruno1}&FEM\\ \hline
            $R_{-1,0}$  &0.01044&0.01164\\
            $R_{0,-1}$  &0.01183&0.01165\\
            $T_{-1,-1}$ &0.06175&0.06299\\ \hline
            $\textcolor{Clarte}{\displaystyle \sum_{(n,m)\in\mathbb{Z}}} \RE \{R_{n,m}\}$ &-&0.10685\\
            $\textcolor{Clarte}{\displaystyle \sum_{(n,m)\in\mathbb{Z}}} \RE \{T_{n,m}\}$ &-&0.89121\\ \hline
            $\textrm{TOTAL}$             &-&0.99806\\ \hline
    \end{tabular}\caption{Energy balance \cite{bruno1}.}\label{tab:result_sinus}
\end{table}
\noindent Note that in order to define this surface, the bi-sinusoid
was first sampled ($15\times15$ points), then converted to a 3D file
format. This sampling can account for the slight differences with
the results obtained using the method of variation of boundaries
developed by Bruno \textit{et al.} (1993).

\begin{figure}[!ht]
    \begin{center}
        \subfloat[] {\includegraphics[width=.45\textwidth,draft=\logic]{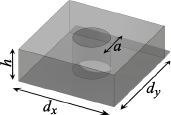}\label{fig:schema_trous}}
            \quad
        \subfloat[] {\includegraphics[width=.45\textwidth,draft=\logic]{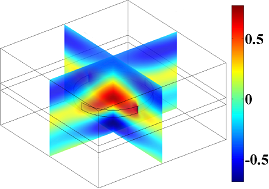}\label{fig:carte_trous}}
    \setlength{\captionmargin}{\mycaptionmargin}
    \caption{Lossy diffractive element with vertical edges (a). $\RE \{E_y\}$ in V/m (b).}\label{fig:trous}
    \end{center}
\end{figure}

\newpage
\paragraph{Circular apertures in a lossy layer}\label{part:trous}
In this example worked out by Schuster \textit{et al.}
\cite{schuster1}, the diffractive element is a circular aperture in
a lossy layer as shown Fig.~\ref{fig:schema_trous} and the grating
parameter  highlighted in Fig.~\ref{fig:schema_3D} are the following:
$\lambda_0 = 500\,$nm, $\varphi_0=\theta_0=0^{\,\circ}$,
$\varepsilon^+ = \varepsilon^{g}=1$, $\varepsilon^{g'}=0.8125 +
5.2500\,i$ and $\varepsilon^-=2.25$.
\begin{table}[!ht]
\centering
    \begin{tabular}{lcccc} \hline
        & \cite{moharam1}&\cite{li3}&\cite{schuster1}&FEM \\ \hline
        $R_{0,0}$&0.24657& 0.24339 &0.24420 & 0.24415\\ \hline
        $\textcolor{Clarte}{\displaystyle \sum_{(n,m)\in\mathbb{Z}}} \RE\{T_{n,m}\}$&$-$& $-$ & $-$ &0.29110\\
        $\textcolor{Clarte}{\displaystyle \sum_{(n,m)\in\mathbb{Z}}} \RE\{R_{n,m}\}$&$-$& $-$ & $-$ &0.26761\\
        $Q$&$-$& $-$ & $-$ &0.44148\\ \hline
        $\textrm{TOTAL}$&$-$ &$-$ &$-$  & 1.00019\\ \hline
    \end{tabular}\caption{Comparison with \cite{moharam1,li3,schuster1} and energy balance.}\label{tab:result_trous}
\end{table}

\begin{figure}[!ht]
    \begin{center}
        \subfloat[] {\includegraphics[width=.35\textwidth,draft=\logic]{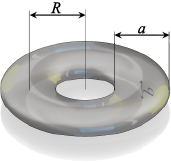}\label{fig:schema_tore}}
            \quad
        \subfloat[] {\includegraphics[width=.35\textwidth,draft=\logic]{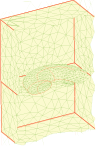}\label{fig:carte_tore}}
    \setlength{\captionmargin}{\mycaptionmargin}
    \caption{Torus parameters (a). Coarse mesh of the computational domain (b).}\label{fig:tore}
    \end{center}
\end{figure}

In this lossy case, results obtained with the FEM show good
agreement with the ones obtained with the FMM \cite{li3}, the
differential method \cite{schuster1,laurent_arnaud} and the RCWA
\cite{moharam1}. Joule losses inside the diffractive element can be
easily calculated, which allows to provide a global energy balance
for this configuration. Finally, the convergence of the value
$R_{0,0}$ \textcolor{Map2}{as a} function of the mesh refinement
will be examined.
\newpage
\paragraph{Lossy tori grating}\label{part:tore}
We finally propose a new test case for crossed-grating numerical
methods. The major difficulty of this case lies both in the non
trivial geometry (see Fig.~\ref{fig:schema_tore}) of the diffractive
object and in the fact that it is made of a material chosen so that
losses are optimal inside it. The grating parameters highlighted in
Fig.~\ref{fig:schema_3D} and Fig.~\ref{fig:schema_tore} are the
following: $\lambda_0=1$, $\varphi_0=\psi_0=0^{\,\circ}$,
$d_x=d_y=0.3\,$, $a= 0.1$, $b= 0.05$, $R=0.15$, $h=500\,$nm,
$\varepsilon^+ =\varepsilon^{g}= 1$, $\varepsilon^{g'}=-21+20\,i$
and $\varepsilon^-=2.25$.

\begin{table}[!ht]
  \centering
  \begin{tabular}{lcc}\\ \hline
        FEM 3D     & $\theta=0^{\circ}$ &$\theta=40^{\circ}$ \\ \hline
        $R_{0,0}$  &         0.36376           & 0.27331\\
        $T_{0,0}$  &         0.32992            & 0.38191\\
        $Q$        &         0.30639             & 0.34476\\ \hline
        $\textrm{TOTAL}$&    1.00007            & 0.99998\\ \hline
  \end{tabular} \caption{Energy balances at normal and oblique incidence.}\label{tab:results_torus}
\end{table}
Tab.~\ref{tab:results_torus} illustrates the independence of our
method towards the geometry of the diffractive element.
$\varepsilon^{g'}$ is chosen so that the skin depth has the same
order of magnitude as $b$, which maximizes losses. Note that energy
balances remain very accurate at normal and oblique incidence, in
spite of both the non-triviality of the geometry and the strong
losses.

\newpage
\subsubsection{Convergence and computation time}\label{part:analyse_valid}
\paragraph{Convergence as a function of mesh refinement}\label{part:converg_3D}
When using modal methods such as the RCWA or the differential
method, based on the calculation of Rayleigh coefficients, a number
proportional to $N_{R}$ have to be be determined \textit{a priori}.
Then, the unknown diffracted field is expanded as a Fourier serie,
injected under this form in Maxwell equations, which annihilates
$x-$ and $y-$dependencies. This leads to a system of coupled partial
differential equations whose coefficients can structured in a matrix
formalism. The resulting matrix is sometimes directly invertible
(RCWA) depending on whether the geometry allows to suppress the
$z-$dependance, which makes this method adapted to diffractive
elements with vertically (or \textcolor{Map2}{decomposed} in
staircase functions) shaped edge. In some other cases, one has to
make the use of integral methods in order to solve the system, as in
the pyramidal case for instance, which leads to the so-called
differential method. The diffracted field map can be deduced from
these coefficients. If the grating configuration only calls for a
few propagative orders and if the field inside the groove region is
not the main information sought for, these two close methods allow
to determine the repartition of the incident energy quickly.
However, if the field inside the groove region is the main piece of
information, it is advisable to calculate many Rayleigh coefficients
corresponding to evanescent waves which increases the computation
time as $(N_R)^3$ or even $(N_R)^4$.

\begin{figure}[!ht]\centering
  \includegraphics[width=.9 \textwidth,draft=\logic]{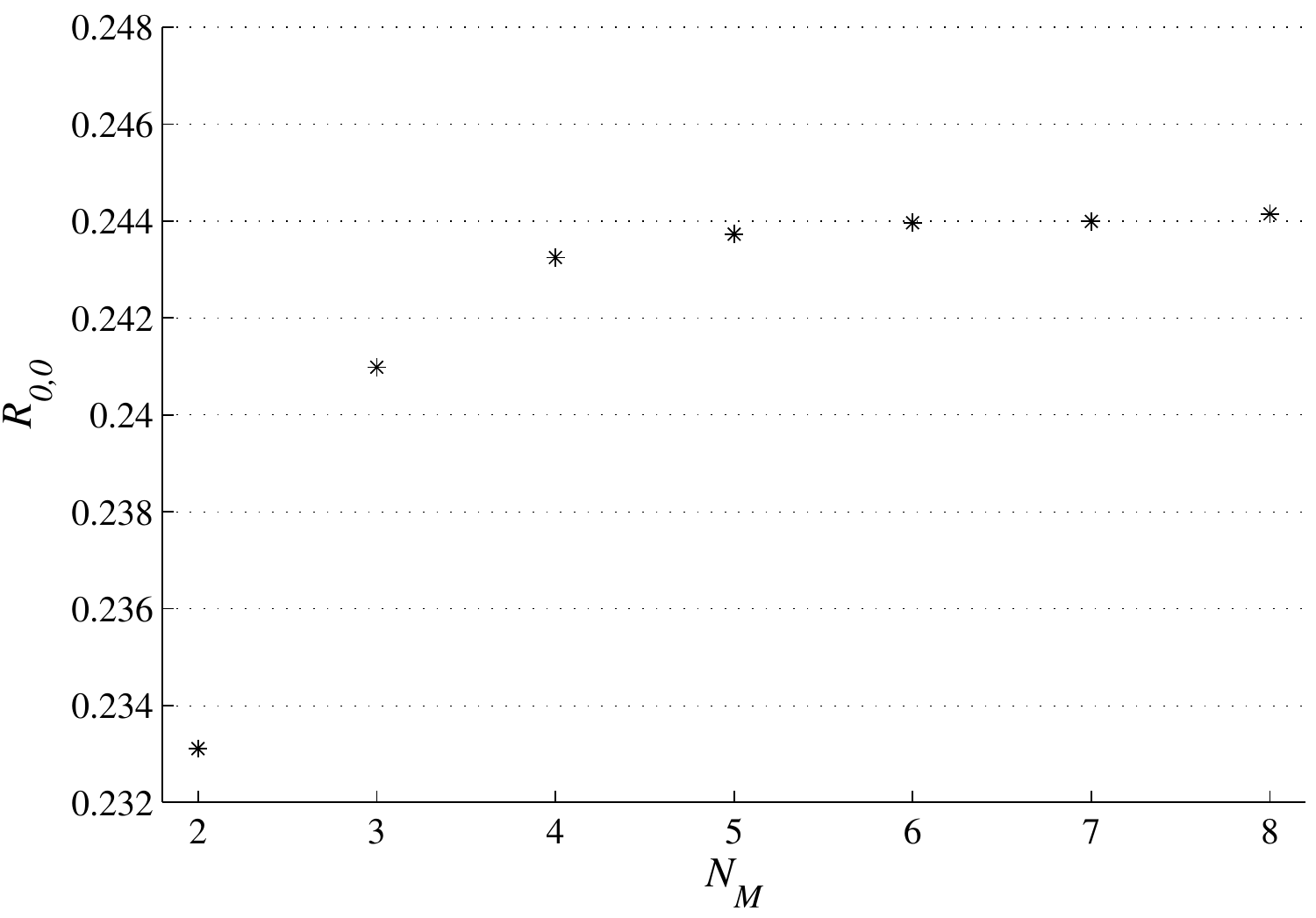}\caption{Convergence of $R_{0,0}$ in function of $N_m$ (circular apertures crossed-grating).} \label{fig:converge}
\end{figure}

FEM relies on the direct calculation of the vectorial components of
the complex field. Rayleigh coefficients are determined \textit{a
posteriori}. The parameter limiting the computation time is the
number of tetrahedral elements along which the computational domain
is split up. We suppose that it is necessary to calculate at least
two or three points (or mesh nodes) per period of the field
($\lambda_0/\sqrt{\RE \{\varepsilon\}}$). Figure~\ref{fig:converge}
shows the convergence of the efficiency $R_{0,0}$ (circular
apertures case, see Fig.~\ref{fig:schema_trous}) \textcolor{Map2}{as
a} function of the mesh refinement characterized by the parameter
$N_M$: The maximum size of each element is set to
$\lambda_0/(N_M\,\sqrt{\RE\{\varepsilon\}})$. 

It is of interest to note that even if $N_M<3$ the FEM still gives
pertinent diffraction efficiencies: $R_{0,0}=\textrm{0.2334}$ for
$N_M = \textrm{1}$ and $R_{0,0}=\textrm{0.2331}$ for
$N_M=\textrm{2}$. \textcolor{Clarte}{The Galerkin method (see
Eq.~(\ref{eq:weak})) corresponds to a minimization of the error
(between the exact solution and the approximation) with respect to a
norm that can be physically interpreted in terms of energy-related
quantities. Therefore, the finite element methods usually provide
energy-related quantities that are more accurate than the local
values of the fields themselves.}
\paragraph{Computation time} \noindent All the calculations were
performed on a server equipped with 8 dual core Itanium1 processors
and 256Go of RAM. Tetrahedral quadratic edge elements were used
together with the direct solver PARDISO. Among different direct
solvers adapted to sparse matrix algebra (UMFPACK, SPOOLES and
PARDISO), PARDISO turned out to be the less time-consuming one as
shown in Tab~\ref{tab:tps_cacl_3D}.

\begin{table}[!ht]
    \begin{center}
        \begin{tabular}{lcc}
            \hline
                Solver & Computation time for 41720 DOF & Computation time for 205198 DOF \\ \hline
                SPOOLES& $15\,$mn$\,32\,$s&$14\,$h$44\,$mn\\
                UMFPACK & $2\,$mn$\,07\,$s&$1\,$h$12\,$mn\\
                PARDISO&$57\,$s&$16\,$mn \\
            \hline 
        \end{tabular}
    \end{center}\caption{Computation time variations from solver to solver.}\label{tab:tps_cacl_3D}
\end{table}

\begin{figure}[!ht]\centering
 \includegraphics[width=.9 \textwidth,draft=\logic]{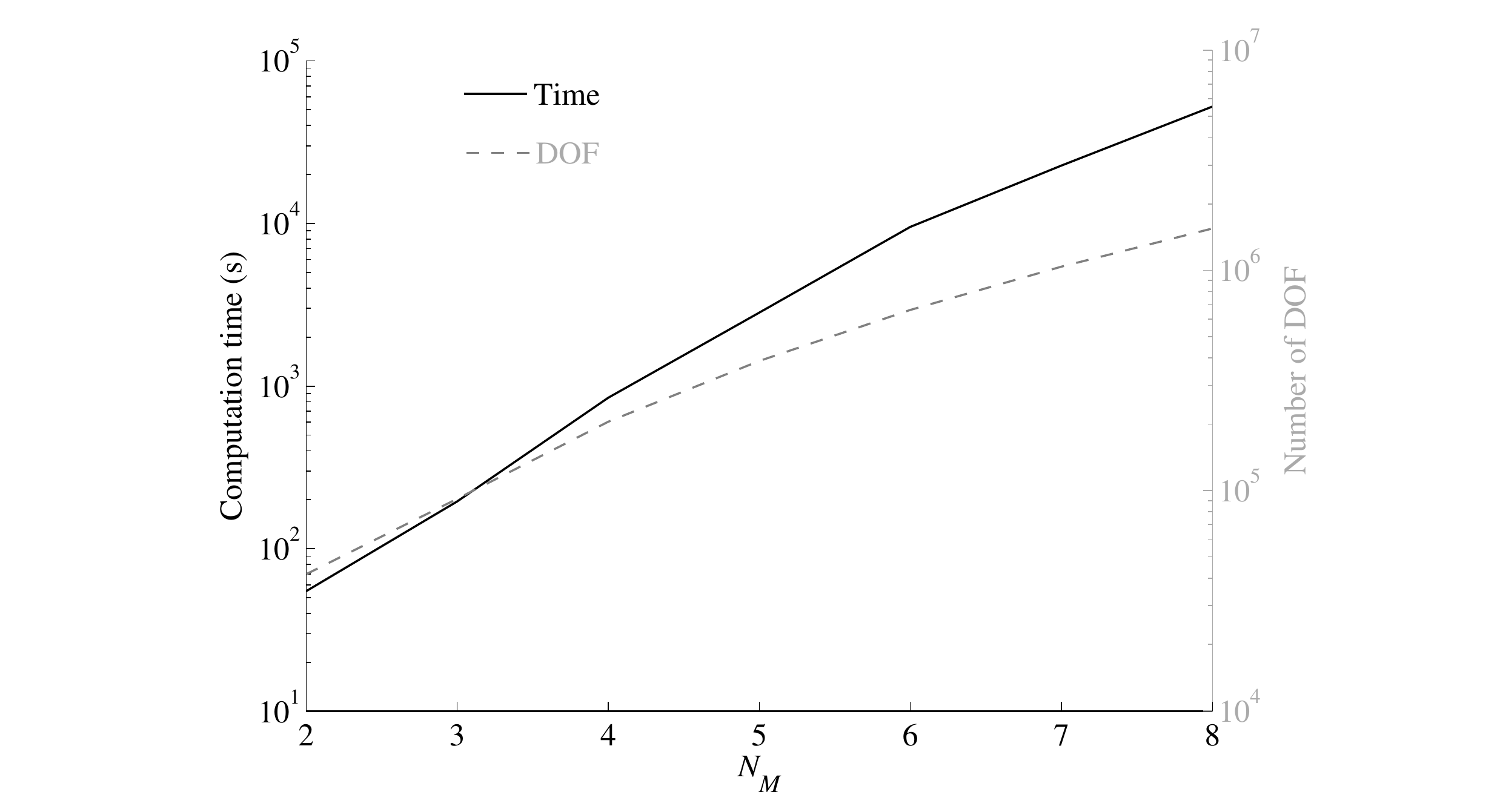}
\caption{Computation time and number of DOF as a function of $N_M$.}
\label{fig:temps_3D}
\end{figure}

Figure~\ref{fig:temps_3D} shows the computation time required to
perform the whole FEM computational process for a system made of a
number of DOF indicated on the right-hand ordinate. It is of
importance to note that for values of $N_M$ lower than 3, the
problem can be solved in less than a minute on a standard laptop
(4Go RAM, $2\times2\,$GHz) with 3 significant digits on the
diffraction efficiencies. This accuracy is more than sufficient in
numerous experimental cases. Furthermore, as far as integrated
values are at stake, relatively coarse meshes ($N_M\approx1$) can be
used trustfully, authorizing fast geometric, spectral or
polarization studies. 

Nowadays, the efficiency of the numerical algorithms for sparse matrix algebra
together with the available power of computers and the fact that the
problem reduces to a basic cell with a size of a small number of
wavelengths make the finite element problem very tractable as proved here.

\section{Concluding remarks}
In this chapter, we demonstrate a general formulation of the FEM allowing
to calculate the diffraction efficiencies from the electromagnetic field
diffracted by arbitrarily shaped gratings embedded in a multilayered stack
lightened by a plane wave of arbitrary incidence and polarization
angle. It relies on a rigorous treatment of the plane wave sources
problem through an equivalent radiation problem with localized
sources. Bloch conditions and a new dedicated PML have been implemented in order to
rigorously truncate the computational domain. 

The principles of the method were discussed in detail for mono-dimensional 
gratings in TE/TM polarization cases (2D or scalar case) in a first part, and for the most general
bi-dimensional or crossed gratings (3D or vector case) in a second part. Note that the very same 
concepts could be applied to the intermediate case of mono-dimensional gratings enlighten by an 
arbitrary incident plane wave (so-called conical case). The reader will find detail about the 
element basis relevant to this case in \cite{nicolet2004modelling}.

The main advantage of this formulation is its complete generality
with respect to the studied geometries and the material properties,
as illustrated with the lossy tori grating non-trivial case. Its
principle remains independent of both the number of diffractive
elements by period and number of stack layers. Its flexibility allowed us to
retrieve with accuracy the few numerical academic examples found in
the literature and established with independent methods.

The remarkable accuracy observed in the case of coarse meshes, makes
it a fast tool for the design and optimization of diffractive
optical components (\textit{e.g.} reflection and transmission filters,
polarizers, beam shapers, pulse compression gratings\dots). The
complete independence of the presented approach towards both the geometry and the
isotropic constituent materials of the diffractive elements makes it
a handy and powerful tool for the study of metamaterials, finite-size photonic
crystals, periodic plasmonic structures\dots
The method described in this chapter has already been successfully applied 
to various problems, from homogenization theory \cite{cuabuz2011homogenization} 
or transformation optics \cite{dupont2009revolution}
to more applied concerns as the modeling of complex CMOS nanophotonic devices 
\cite{demesy2009finite} or ultra-thin new generation solar cells \cite{demesy2012solar}.
\bla

\newpage
\setcounter{section}{0}
\renewcommand{\thesection}{\thechapter.\Alph{section}}

\section{APPENDIX}\label{annexe_multi}
This appendix is dedicated to the determination of the vector
electric field in a dielectric stack enlightened by a plane wave of
arbitrary polarization and incidence angle. This calculation,
abundantly treated in the 2D scalar case, is generally not presented
in the literature since, as far as isotropic cases are concerned, it
is possible to project the general vectorial case on the two
reference TE and TM cases. However, the presented formulation can be
extended to a fully anisotropic case for which this TE/TM decoupling
is no longer valid and the three components of the field have to be
calculated as follows.

Let us consider the \textit{ancillary problem} mentioned in
Sec.~\ref{part:radiation_pb}, \textit{i.e.} a dielectric stack made of
$N$ homogeneous, isotropic, lossy layers characterized by there
relative permittivity denoted $\varepsilon^j$ and their thickness
$e_j$. This stack is deposited on a homogeneous, isotropic, possibly
lossy substrate characterized by its relative permittivity denoted
$\varepsilon^{N+1}=\varepsilon^-$. The superstrate is air and its
relative permittivity is denoted $\varepsilon^+=1$. Finally, we
denote by $z_j$ the altitude of the interface between the $j^{th}$
and $j+1^{th}$ layers. The restriction of the incident field
$\bE^{\mathrm{inc}}$ to the superstrate region is denoted $\bE_0$.
The problem amounts to looking for ($\bE_1$,$\bH_1$) satisfying
Maxwell equations in harmonic regime (see
Eqs.~(\ref{eq:MaxwellrotE2},\ref{eq:MaxwellrotH2})).

\subsection*{Across the interface $z=z_j$} By projection on the
main axis of the vectorial Helmholtz propagation equation
(Eq.~(\ref{eq:defLvect})), the total electric field inside the $j^{th}$
layer can be written as the sum of a propagative and a
counter-propagative plane waves:
\begin{equation}\label{eq:defE}
\bE_1(x,y,z) = \left[ \begin{array}{l}
E_1^{x,j,+}\\E_1^{y,j,+}\\E_1^{z,j,+}
    \end{array} \right]
    \exp\left(j\,(\alpha_0\,x+\beta_0\,y+\gamma_j\,z)\right)+
    \left[ \begin{array}{l} E_1^{x,j,-}\\E_1^{y,j,-}\\E_1^{z,j,-}
    \end{array} \right] \exp\left(j\,(\alpha_0\,x+\beta_0\,y-\gamma_j\,z)\right)
\end{equation}
where
\begin{equation}
\gamma_j^2 = k_j^2-\alpha_0^2-\beta_0^2
\end{equation}
What follows consists in writing the continuity of the tangential
components of $(\bE_1,\bH_1)$ across the interface $z=z_j$,
\textit{i.e.} the continuity of the vector field $\Psi$ defined by:
\begin{equation}
    \Psi = \left[ \begin{array}{l} E_1^x\\E_1^y\\i\,H_1^x\\i\,H_1^y
    \end{array} \right] .
\end{equation}
The continuity of $\Psi$ along $Oz$ together with its analytical
expression inside the $j^{th}$ and $j+1^{th}$ layers allows to
establish a recurrence relation for the interface $z=z_j$.

Then, by projection of
Eqs.~(\ref{eq:MaxwellrotE2},\ref{eq:MaxwellrotH2}) on $Ox$,$Oy$ and
$Oz$:
\begin{equation}\label{eq:annexe_cm_vect_maxwell1}
\left[ \begin{array}{l}
    i\,\beta_0\,H_1^z-\frac{\partial H_1^y}{dz}\\
    \frac{\partial H_1^x}{dz}-i\,\alpha_0\,H_1^z\\
    i\,\alpha_0\,H_1^y-i\,\beta_0\,H_1^x
\end{array} \right] =
-i\, \omega\, \varepsilon \left[ \begin{array}{l}
    E_1^x\\
    E_1^y\\
    E_1^z
\end{array} \right]
\end{equation}
and
\begin{equation}\label{eq:annexe_cm_vect_maxwell2}
\left[ \begin{array}{l}
    i\,\beta_0\,E_1^z-\frac{\partial E_1^y}{\partial z}\\
    \frac{\partial E_1^x}{\partial z}-i\,\alpha_0\,E_1^z\\
    i\,\alpha_0\,E_1^y-i\,\beta_0\,E_1^x
\end{array} \right] =
i\, \omega\, \mu \left[ \begin{array}{l}
    H_1^x\\
    H_1^y\\
    H_1^z
\end{array} \right] .
\end{equation}
Consequently, tangential components of $\bH_1$ can be expressed in
function of tangential components of $\bE_1$:
\begin{equation}
\underbrace{    \begin{bmatrix}
        \omega\,\mu&0&\beta_0\\
        0&\omega\,\mu&-\alpha_0\\
        -\beta_0&\alpha_0&-\omega\,\varepsilon
    \end{bmatrix}\,}_{B}
    \begin{bmatrix}
        i\,H_1^x\\
        i\,H_1^y\\
        i\,H_1^z
    \end{bmatrix}
    =
    \begin{bmatrix}
        \frac{\partial E_1^y}{dz}\\
        -\frac{\partial E_1^x}{dz}\\
        0
    \end{bmatrix} .
\end{equation}
By noticing the invariance and linearity of the problem along $Ox$
and $Oy$, the following notations are adopted:
\begin{equation}
    \left\{ \begin{array}{l}
    U_x^{j,\pm} =E_1^{x,j,\pm}\,\exp(\pm\,i\,\gamma_j\,z)\\
    U_y^{j,\pm} =E_1^{y,j,\pm}\,\exp(\pm\,i\,\gamma_j\,z) \end{array}
    \right.
\end{equation} and
\begin{equation}
    \Phi_j = \left[ \begin{array}{l}
    U_x^{+,j}\\U_x^{-,j}\\U_y^{+,j}\\U_y^{-,j}
        \end{array} \right] .
\end{equation}
Thanks to Eq.~(\ref{eq:defE}) and
Eq.~(\ref{eq:annexe_cm_vect_maxwell2}) and letting $M=B^{-1}$, it
comes for the $j^{th}$ layer:
\begin{equation}
\Psi(x,y,z) = \exp(i(\alpha_0\,x+\beta_0\,y))\,\underbrace{
\begin{bmatrix}
        1&1&0&0\\
        0&0&1&1\\
        \gamma_j\,M^j_{12}&-\gamma_j\,M^j_{12}&-\gamma_j\,M^j_{11}&\gamma_j\,M^j_{11}\\
        \gamma_j\,M^j_{22}&-\gamma_j\,M^j_{22}&-\gamma_j\,M^j_{21}&\gamma_j\,M^j_{21}\\
    \end{bmatrix}}_{\Pi_j}\,\left[ \begin{array}{l}
U_x^{+,j}\\U_x^{-,j}\\U_y^{+,j}\\U_y^{-,j}
    \end{array} \right] .
\end{equation}
Finally, the continuity of $\Psi$ at the interface $z=z_j$ leads to:
\begin{equation}\label{eq:passage}
    \Phi_{j+1}(z_j) = \Pi_{j+1}^{-1}\,\Pi_j\,\Phi_{j}(z_j) .
\end{equation}
Normal components can be deduced using
Eqs.~(\ref{eq:annexe_cm_vect_maxwell1},\ref{eq:annexe_cm_vect_maxwell2}).
\subsection*{Traveling inside the $j+1^{th}$ layer} Using
Eq.~(\ref{eq:defE}), a simple phase shift allows to travel from
$z=z_{j}$ to $z=z_{j+1}=z_j-e_{j+1}$:
\begin{equation}\label{eq:transport}
\Phi_{j+1}(z_{j+1}) = \underbrace{
\begin{bmatrix}
        \exp(-i\,\gamma_{j+1}\,e_{j+1})&0&0&0\\
        0&\exp(+i\,\gamma_{j+1}\,e_{j+1})&0&0\\
        0&0&\exp(-i\,\gamma_{j+1}\,e_{j+1})&0\\
        0&0&0&\exp(+i\,\gamma_{j+1}\,e_{j+1})\\
    \end{bmatrix}}_{T_{j+1}}\,\Phi_{j+1}(z_{j})
\end{equation}
Thanks to Eq.~(\ref{eq:transport}) and Eq.~(\ref{eq:passage}), a
recurrence relation can be formulated for the analytical expression
of $\bE_1$ in each layer:
\begin{equation}\label{eq:recurrence}
\Phi_{j+1}(z_{j+1}) =
T_{j+1}\,\Pi_{j+1}^{-1}\,\Pi_j\,\Phi_{j}(z_{j})
\end{equation}
\subsection*{Reflection and transmission coefficients} The last step
consists in the determination of the first term $\Phi_0$, which is
not entirely known, since the problem definition only specifies
$U_x^{0,+}$ and $U_y^{0,+}$, imposed by the incident field $\bE_0$.
Let us make the use of the OWC hypothesis verified by $\bE_1^d$ (see
Eq.~(\ref{eq:defL1_vect})). This hypothesis directly translates the
fact that none of the components of $\bE_1^d$ can either be
traveling down in the superstrate or up in the substrate:
$U_y^{N+1,-}=U_x^{N+1,-}=0$. Therefore, the four unknowns
$U_x^{0,-}$, $U_y^{0,-}$,$U_y^{N+1,+}$ and $U_x^{N+1,+}$,
\textit{i.e.} transverse components of the vector fields reflected and
transmitted by the stack, verify the following equation system:
\begin{equation}
    \Phi_{N+1}(z_{N}) =(\Pi_{N+1})^{-1}\,\Pi_N
    \prod_{j=0}^{N-1}T_{N-j}\,(\Pi_{N-j})^{-1}\,\Pi_{N-j-1}\,
    \Phi_0(z_0)
\end{equation}
This allows to extend the definition of transmission and reflection
widely used in the scalar case. Finally, $\Phi_{N+1}$ is entirely
defined. Making the use of the recurrence relation of
Eq.~(\ref{eq:recurrence}) and of Eq.~(\ref{eq:defE}) leads to an
analytical expression for $\bE_1^d$ in each layer.

\let\cleardoublepage\clearpage
\renewcommand\bibname{\normalsize{\hspace{12 pt}References:}}

\end{document}